\documentclass{aa}
\usepackage{psfig}
\usepackage{lscape}
\usepackage{longtable}
\usepackage{amsmath}
\usepackage{amssymb}
\usepackage[figuresleft]{rotating}

\newcommand{\MC}{\multicolumn}
\newcommand{\kms}{km~s$^\mathrm{-1}$}
\newcommand{\sunn}{$_{\odot}$}

\newcounter{qub}
\newcommand{\qq}{\addtocounter{qub}{1}\arabic{qub}}

\begin{document}

\title{
The Hamburg/SAO Survey for low metallicity blue compact/\ion{H}{ii} galaxies
(HSS--LM). }
\subtitle{I. The first list of 46 strong-lined galaxies}

\author{A.V. Ugryumov\inst{1,4}  \and
D. Engels\inst{2} \and
S.A. Pustilnik\inst{1,4} \and
A.Y. Kniazev\inst{1,3,4} \and
A.G. Pramskij\inst{1,4} \and
H.-J. Hagen\inst{2}
}

\offprints{A.V.Ugryumov, \email{and@sao.ru}}

\institute{
Special Astrophysical Observatory RAS, Nizhnij Arkhyz,
Karachai-Circassia, 369167 Russia
\and Hamburger Sternwarte, Gojenbergsweg 112, 21029 Hamburg, Germany
\and Max Planck Institut f\"{u}r Astronomie, K\"{o}nigstuhl 17, D-69117,
Heidelberg, Germany
\and Isaac Newton Institute of Chile, SAO Branch
}

\date{Received 28 January, 2002 / Accepted 10 September, 2002}

\abstract{
We present the description and the first results\thanks{Tables 2 and 3 are
only available in electronic form at CDS via anonymous ftp to
cdsarc.u-strasbg.fr (130.79.128.5) or via http://cdsweb.u-strasbg.fr/Abstract.html.
Figures A1--A6 are only  available in electronic form at Edpsciences
(http://www.edpsciences.org)} of a new project devoted
to the search for extremely metal-deficient blue compact/\ion{H}{ii}-galaxies
(BCGs) and to the creation of a well selected large BCG sample with strong
emission lines. Such galaxies should be suitable for reliable determination
of their oxygen abundance through the measurement of the faint
[\ion{O}{iii}]\,$\lambda$\,4363~\AA\ line. The goals of the project are
two-fold:
a) to discover a significant number of new extremely metal-poor galaxies
($Z \lesssim$ 1/20~$Z$\sunn), and
b) to study the metallicity distribution of local BCGs. Selection
of candidates for follow-up slit spectroscopy is performed on the database
of objective prism spectra of the Hamburg Quasar Survey. The sky region is
limited by $\delta \geq 0$\degr\ and $b^{II} \leq -30$\degr. In this paper
we present the results of the follow-up spectroscopy conducted with
the Russian 6\,m telescope.
The list of observed candidates contained 52 objects, of which 46 were
confirmed as strong-lined  BCGs ($EW$([\ion{O}{iii}]\,$\lambda$\,5007)
$\ge$ 100~\AA). The remaining five lower excitation ELGs include three BCGs,
and two galaxies classified as SBN (Starburst Nucleus) and DANS (Dwarf
Amorphous Nucleus Starburst).
One object is identified as a quasar with a strong Ly$\alpha$ emission line
near $\lambda$\,5000~\AA\ (z~$\sim 3$). We provide a list with coordinates,
measured radial velocities, $B$-magnitudes, equivalent widths
$EW$([\ion{O}{iii}]\,$\lambda$\,5007) and $EW$(H$\beta$) and for the 46
strong-lined BCGs the derived oxygen abundances 12+$\log$(O/H).
The abundances range between 7.42 and 8.4 (corresponding to metallicities
between 1/30 and 1/3~$Z$\sunn). The sample contains four galaxies with
$Z \lesssim$ 1/20~$Z$\sunn, of which three are new discoveries.
This demonstrates the high efficiency of the new project to find extremely
metal-deficient galaxies. The radial velocities of the strong--lined ELGs
range between 500 and 19000~\kms\ with a median value of $\sim$~6400~\kms.
The typical $B$-magnitudes of the galaxies presented are $17\fm0-18\fm0$.
}

\authorrunning{Ugryumov et al.}

\titlerunning{The Hamburg/SAO Survey for low metallicity
BCG. I.}

\maketitle

\section{Introduction}

Blue compact/\ion{H}{ii} galaxies (BCGs) are low-mass gas-rich objects which
are currently undergoing an episode of enhanced star formation.
Such episodes are usually
recognized by a strong emission-line spectrum of \ion{H}{ii} type.
Their duration is relatively short: strong emission lines are detectable on
timescales from ten to a hundred Myr.
These episodes of intense star formation are often called starbursts.
BCGs have been intensively studied since the seminal work by Sargent \& Searle
(\cite{Sargent70}), and has been especially actively during the last decade.
The list of publications on this subject is very long, and we list below
only those
studies dealing with observations and/or analysis of sufficiently large
samples. These papers, among others, are Campos-Aguilar et al.
(\cite{Campos93}), Masegosa et al. (\cite{Masegosa94}), Izotov et al.
(\cite{Izotov94}), Thuan et al. (\cite{Thuan95}),  Salzer et al.
(\cite{Salzer95}), Pustilnik et al. (\cite{PULTG}), Taylor et al.
(\cite{Taylor95}), Vilchez (\cite{Vilchez95}), Stasinska \& Leitherer
(\cite{SL96}), Telles \& Terlevich (\cite{Telles97}), van Zee et al.
(\cite{Zee98a}), Schaerer et al. (\cite{Schaerer97}, \cite{Schaerer99}),
Izotov \& Thuan~(\cite{Izotov99}), Bergvall et al. (\cite{Bergvall99}),
Thuan et al. (\cite{TLMP99}), Kunth \& \"Ostlin (\cite{KO}),
Pustilnik et al. (\cite{BCG_ENV}), and Ugryumov et al.~(\cite{Ugryumov01},
and references therein).

To date more than thousand galaxies of this type are known. Some of them had
been found in early studies as Zwicky compact galaxies (Zwicky
\cite{Zwicky}) or Haro blue galaxies (Haro \cite{Haro56}). However, the
great majority of BCGs were picked up by objective prism surveys such as the
First and Second Byurakan (FBS, SBS) (Markarian~\cite{Markarian67};
Markarian et al.~\cite{Markarian83};
Stepanian~\cite{Stepanian94}), the University of Michigan (Salzer et al.
\cite{Salzer89b}), the Tololo (Terlevich et al.~\cite{Terlevich91}), the Case
(Pesch et al.~\cite{Pesch95}; Salzer et al.~\cite{Salzer95}; Ugryumov et al.
\cite{Ugryumov98}), the Heidelberg void (Popescu et al. \cite{Popescu98})
and the Hamburg/SAO survey (Ugryumov et al.~\cite{Ugryumov01}, and references
therein).

\begin{figure}
\centering
\psfig{figure=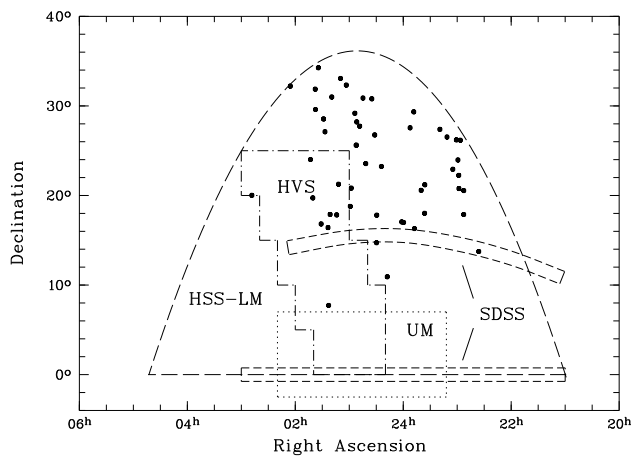,width=9cm,angle=-90}
\caption[]{\label{Fig1}
 The sky area covered by the HSS--LM is determined by $b^{II}~\le~-30$\degr\
 and $\delta \ge$ 0\degr. The positions of ELGs presented in this paper
 are shown with  filled circles.  The zones of three other surveys:
 University of Michigan (UM), Heidelberg void survey (HVS) and Sloan Digital
 Sky Survey (SDSS) are shown superimposed on the HSS--LM region (see
 references in the text).
}
\end{figure}

Like other low-mass galaxies, BCGs have
metallicities $Z$ many times lower than those of normal bright galaxies. Their
characteristic range of metallicity $Z$ is $\sim$~($1/15-1/3$)~$Z$\sunn\
(e.g., Terlevich et al.~\cite{Terlevich91}, Izotov et al. \cite{Izotov93}),
while
very few of them show $Z$ as low as ($1/50-1/20$)~$Z$\sunn. All well-studied
BCGs with characteristic metallicities were found to be old galaxies, which
evolve slower than their massive counterparts. Instead of having a lower gas
consumption rate, it was discussed many years galactic winds were
responsible for a significant loss of heavy elements
(Mac Low \& Ferrara \cite{MLF99}); however more recent models
(Silich \& Tenorio-Tagle \cite{STT01}, Legrand et al. \cite{Legrand02})
indicate that for the majority of BCGs a significant loss of metals is rather
improbable.
In contrast to all typical BCGs, a few well studied BCGs with an extreme
deficit of heavy elements
show properties of young galaxies, undergoing their first SF episode. In
particular, they have gas-mass fractions
(that is, gas mass relative to all visible baryon mass) of the order $0.9-0.99$
(e.g., Salzer et al. \cite{Salzer91},
Kniazev et al. \cite{Kniazev00b}, Pustilnik et al. \cite{PBTLI}) and in the
outermost parts of their low-surface brightness disks show very blue colours,
consistent with ages of the ``old'' stellar population less than $100-200$~Myr.

One of the most often discussed candidate objects for low age is I~Zw~18
with $Z = 1/50~Z$\sunn, discovered by Searle \& Sargent (\cite{Searle72}).
The most recent data are summarized by Izotov \& Thuan (\cite{Izotov99}),
Izotov et al. (\cite{Izotov01}), van Zee et al. (\cite{Zee98b}),
Aloisi et al. (\cite{Aloisi99}), \"Ostlin (\cite{Ostlin00}), and
Papaderos et al. (\cite{Papaderos02}).

While the first two papers forward arguments for a maximum age of
less than 200~Myr, Aloisi et al. (\cite{Aloisi99}) and \"Ostlin (\cite{Ostlin00})
suggest an age of at least 1~Gyr. The age of the extremely metal-poor
galaxies remains therefore a matter of debate.

Several other well-known objects are SBS~0335--052 (E and W) with
$Z = 1/42~Z$\sunn\ and 1/50~$Z$\sunn\ (e.g., Izotov et al. \cite{Izotov97};
Papaderos et al. \cite{Papaderos98}; Lipovetsky et al. \cite{Lipovetsky99};
Pustilnik et al. \cite{PBTLI}, \"Ostlin \& Kunth \cite{OK01}),
HS~0822+3542 with $Z = 1/30~Z$\sunn\
(e.g., Kniazev et al. \cite{Kniazev00b}; Pustilnik et al. \cite{LSBD};
Izotov et al. \cite{Izotov02}), Tol~1227--273 with $Z = 1/28~Z$\sunn\
(e.g., Fricke et al. \cite{Fricke01}) and CG~389 (1415+437) with
$Z = 1/22~Z$\sunn\  (e.g., Thuan et al. \cite{Thuan99}).

Most of the evidences is still indirect and in many cases controversial,
but the study of these rare galaxies clearly stimulates our understanding of
the earliest periods of galaxy evolution. Probably, these extremely
metal-deficient BCGs are the best approximation of low-mass young
galaxies expected to form commonly at redshifts $3 < z < 5$.

Despite 30 years of observations and more than a thousand known BCG galaxies,
another important aspect of BCG studies remains unclear.
What is the true metallicity distribution of BCGs and their progenitors?
This has important cosmological implication, because the knowledge of the
contemporary  $Z$ distribution allows one to pose question
regarding their chemical evolution during the last few Gyr.

The results of BCG morphological studies indicate that they do not comprise
a homogeneous group, but rather are a mixture of various types
(e.g. Kunth et al. \cite{Kunth88}; Doublier et al. \cite{Doublier97}).
This implies that BCG progenitors probably are not a rare type of gas-rich
low-mass galaxy (see, however, Salzer \& Norton \cite{SN99}).
On the other hand, the recent results by Mouri \&
Taniguchi (\cite{Mouri00}), based on far--IR tracers of recent SF,
indicate that up to $\sim(30-40)$\% of the galaxies in their magnitude-limited
sample within the galactic neighborhood have experienced SF activity during
the last 100 Myr.
The evidence that such a large fraction of galaxies experience episodes of
significantly enhanced SF emphasizes the importance of BCG studies.
In particular, the study of galaxies in the BCG phase allows us to more
easily understand some of their progenitor parameters (such as $Z$), which
are difficult to determine in more quiet periods of galaxy evolution.

Based on a large well-selected sample of BCGs and using the model
predictions for the evolution of observable parameters, such as
$EW$(H$\beta$), we hope to advance in establishing the true metallicity
distribution of BCGs. This would be a major step to obtain the true
$Z$-distribution of the BCG parent population -- gas-rich low-mass galaxies.

These ideas are the motivation of our new project, ``The Hamburg/SAO
survey for low metallicity BCG/\ion{H}{ii} galaxies''.
The sky region covered is limited by the southern galactic hemisphere
north of $\delta=0$\degr\ and $b^{II}\leq -$30\degr. This large section of
the sky is not well covered by spectroscopic surveys of sufficient depth.
Besides the First Byurakan Survey (Markarian~\cite{Markarian67}) dealing
with relatively bright galaxies, a 9.5\degr wide strip of the University
of Michigan (UM) Survey (MacAlpine et al.~\cite{McAlpine77};
Salzer et al.~\cite{Salzer89b} and references therein) and of several fields
of the Heidelberg void survey (HVS) (Hopp \& Kuhn \cite{Hopp95}; Popescu et al.
\cite{Popescu96}) there are no large deep surveys in this region.
It will be covered in the near future by two 1.5\degr\ wide strips
of the SDSS,
having good spectroscopy for all galaxies brighter than $B \sim 18\fm5$
(York et al. ~\cite{York2000}), and in the more distant future this
region will be covered by the 6dF-z project (Mamon \cite{Mamon99}).
The latter will obtain spectra only for relatively bright galaxies selected
in the near-IR corresponding to $B \lesssim 16\fm0$ for blue galaxies and
hence will not compete with our significantly deeper survey.
Therefore, most of the results obtained in the frame of this
new survey will be valuable material for studies of strong-lined
BCGs/\ion{H}{ii} galaxies in this region.

This project has several features in common with the
earlier ``Hamburg/SAO survey for emission-line galaxies'' (HSS for ELGs),
presented in a series of papers by Ugryumov et al. (\cite{Ugryumov99})~(I),
Pustilnik et al. (\cite{Pustilnik99})~(II), Hopp et al.
(\cite{Hopp00})~(III), Kniazev et al. (\cite{HSS_4})~(IV), Ugryumov et al.
(\cite{Ugryumov01})~(V) and Pustilnik et al. (\cite{PEM02})~(VI).

The new project we will abbreviate as ``HSS--LM'' (LM --- for low metallicity)
to distinguish it from the ``HSS'' (for ELGs). The selection procedure used in
the HSS is described in detail by Ugryumov et al. (\cite{Ugryumov99}).
The main difference introduced in the HSS--LM is a more strict criterion
on the strength of emission-line features in the full-resolution objective
prism spectra.
We select for follow-up long-slit spectroscopy only candidate
galaxies with the strongest emission feature near $\lambda$\,5000\,\AA,
which, according to our previous experience, corresponds to the
([\ion{O}{iii}]\,$\lambda$\,5007) line with equivalent widths
$EW > 150-200$~\AA.
Such galaxies are quite rare and their prism spectra are prominent
enough not to miss them.
Our slit spectroscopy justifies the correctness of the elaborated
criterion. The great majority of the galaxies observed indeed have
$EW$([\ion{O}{iii}]\,$\lambda$\,5007) $>$ 150~\AA, and the fraction
of galaxies with $EW$([\ion{O}{iii}]\,$\lambda$\,5007)
$<$ 100~\AA\ is low (see Fig.~\ref{Fig3}).

The main advantage of this sample is that for most of these faint candidates
the 6\,m telescope spectra with moderate integration times ($\sim20$ min.)
give a well detected [O{\sc iii}]\,$\lambda$\,4363\,\AA\ line, allowing the
reliable determination
of oxygen abundance by the standard method (Pagel et al. \cite{Pagel92}).

In the context of an instantaneous SF burst the imposed selection criterion
selects galaxies with very young SF bursts: typically younger than
$5-10$ Myr (Schaerer \& Vacca \cite{SV98}; Leitherer et al.
\cite{Starburst99}). However, recent results
indicate that quite often SF in BCGs proceeds in a more complicated mode,
similar to a propagating wave of SF along the galaxy body
(e.g., Zenina et al.~\cite{Zenina97}; Thuan et al.~\cite{Thuan99}
among others). Probably in this case, lines with large
$EW$([\ion{O}{iii}]\,$\lambda$\,5007) come from regions close to
the front of such a propagating SF wave.

In this paper we present the general outline of the project, give the details
of the selection procedure and present the first list of galaxies observed
spectroscopically.
In Section~2 the selection criteria are discussed. In Section~3 the
follow-up spectral observations and the data reduction are described.
We present some results of the spectrophotometric observations and a
brief summary
in Section~4. The efficiency of this survey in detecting strong-lined ELGs
and its perspectives are discussed and the respective conclusions are drawn in
Section~5. Throughout the
paper a Hubble constant of H$_{0} = 75$~\kms\mbox{Mpc}$^{-1}$ is used.

\section{Sample selection}
\label{sample}

The objects presented in this paper were selected from the digitized
objective prism plates of the Hamburg Quasar Survey (Hagen et al.
\cite{Hagen95}).

\subsection{Overview of the HSS selection}

The original selection procedure used in the HSS was described
in detail by Ugryumov et al. (\cite{Ugryumov99}), and
some improvements were added in Kniazev et al. (\cite{HSS_4}).
The HSS selection procedure picked up efficiently \ion{H}{ii}-galaxies with
$EW$([\ion{O}{iii}]\,$\lambda$\,5007) $>$ 50~\AA.
It was optimized to deal with a large number of spectra ($30\,000-50\,000$ per
plate) and aimed to reach a balance between a good completeness and a
sufficiently high surface density in order to allow a
study of their large-scale spatial distribution.

\begin{figure}
\centering
\psfig{figure=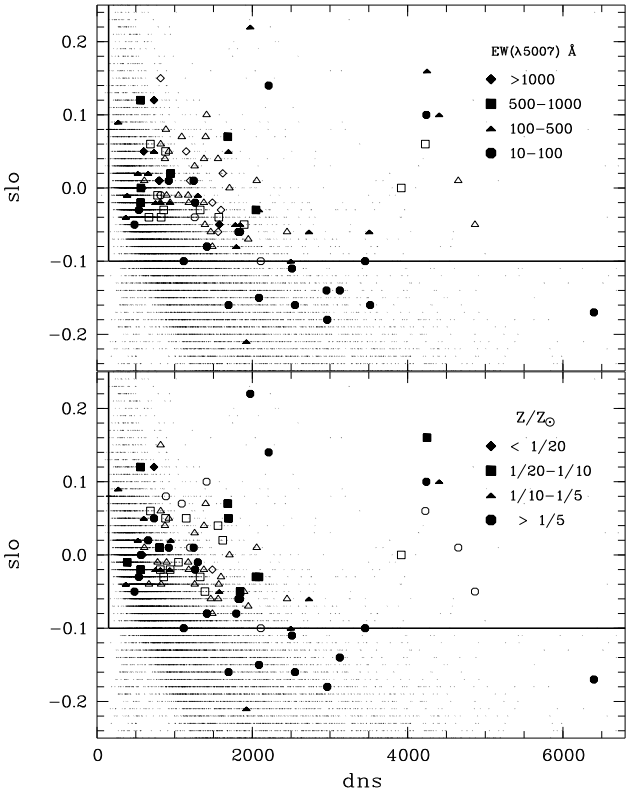,width=9.7cm,angle=0}
\vspace{-2.5cm}
\caption[]{\label{Fig2}
{\it Top:}
 Positions of known galaxies from the training sample of BCGs from the SBS
 (filled symbols) and of BCGs from this paper (open symbols) in the
 plane of the two parameters characterizing the digitized low resolution
 prism spectra  (LRS):
 the integral density (brightness) (dns) versus the slope of spectra (slo),
 This is to be compared to the parameters of objects of all types, as
 recorded from the  HQS plates (dots plotted with the step of 0.01 on dns).
 Different symbols denote objects with different
 $EW$([\ion{O}{iii}]\,$\lambda$\,5007)
 as indicated in the upper-right corner.\\
 {\it Bottom:} Same as above, but for galaxies with known metallicity.
 Different symbols denote galaxies with different metallicities.
 The rectangle indicates the final selection box (see text for details).
}
\end{figure}

This procedure consisted of several steps. The first step was
performed on the prism spectra digitized with low resolution (LRS). It
allowed us to discard more than 2/3 of all spectra that fall
outside the predetermined region in the two-parameter (``slope'', ``density'')
space for LRS spectra. In the next step all preselected LRS were
checked on a graphic screen to pick up the candidates with hints
[\ion{O}{iii}]\,$\lambda$\,5007~\AA\ emission in order to reduce the number
of candidates down to $3000-5000$. This number of LRS was sufficiently small
to rescan them with the Hamburg PDS~1010\,G microdensitometer in a
high-resolution mode in order to get a
confident selection of ELGs for follow-up spectroscopy.

\subsection{Selection of HSS--LM candidates from the HQS database}

In the case of the HSS--LM the selection procedure adopted for the HSS
was changed in order to speed up the selection and to increase the
efficiency of the follow-up spectroscopy. The reason for this change is the
different goals of the HSS--LM in comparison to those of the HSS.
The former is intended
to collect in a reasonable time a large sample of strong-lined ELGs with a
well-defined selection function and $EW$([\ion{O}{iii}]\,$\lambda$\,5007)
$>$ $150-200$~\AA, among which we expect to find about half a dozen  new
extremely metal-deficient BCGs.

%
%
\begin{table*}
\begin{center}
\caption{\label{Tab1} Journal of observations at the SAO 6\,m telescope}
\begin{tabular}{ccrllccc} \\ \hline
\MC{1}{c}{ Run } &
\MC{3}{c}{ Date } &
\MC{1}{c}{ Instrument } &
\MC{1}{c}{ Grating } &
\MC{1}{c}{ Wavelength } &
\MC{1}{c}{ Dispersion } \\

\MC{1}{c}{ No } & & & &  &
\MC{1}{c}{ [grooves\,mm$^{-1}$] } &
\MC{1}{c}{ Range [\AA] } &
\MC{1}{c}{ [\AA\,pixel$^{-1}$] } \\

\MC{1}{c}{ (1) } &
\MC{3}{c}{ (2) } &
\MC{1}{c}{ (3) } &
\MC{1}{c}{ (4) } &
\MC{1}{c}{ (5) } &
\MC{1}{c}{ (6) } \\
\hline
\\[-0.3cm]
\qq& 12--15 & Jul & 1999  & CCD, LSS     & 325 & 3600--7800 & 4.6  \\
\qq&     12 & Aug & 1999  & CCD, LSS     & 650 & 3700--6100 & 2.4  \\
\qq& 05--06 & Sep & 1999  & CCD, LSS     & 650 & 3700--6100 & 2.4  \\
\qq& 01, 03 & Nov & 1999  & CCD, LSS     & 325 & 3600--7800 & 4.6  \\
\qq& 27--30 & Jul & 2000  & CCD, LSS     & 650 & 3700--6100 & 2.4  \\
\qq& 03--06 & Oct & 2000  & CCD, LSS     & 650 & 3700--6100 & 2.4  \\
\hline \\[--0.2cm]
\end{tabular}
\end{center}
\end{table*}

The choice of this limit is related to three facts.
First, all but one known BCGs with $Z \le 1/20~Z_{\odot}$ have
$EW$([\ion{O}{iii}]\,$\lambda$\,5007) $>$ 200~\AA. Second, in
follow-up medium S/N spectra of faint galaxies, having the
[\ion{O}{iii}]-line weaker than the limit indicated, the line
[\ion{O}{iii}]\,$\lambda$\,4363~\AA\ (which is 120 to $\sim 40$ times
weaker than
[\ion{O}{iii}]\,$\lambda$\,5007~\AA\ for $T_\mathrm{e}$ in the range 12\,000
to 20\,000~K, Pagel et al. \cite{Pagel92})
will be barely measurable, preventing a reliable determination of O/H.
And last, as shown by models of Schaerer \& Vacca (\cite{SV98}), the decrease
expected for the H$\beta$ equivalent widths after an instantaneous starburst
is slower for lower metallicity models. For example, for $Z = 1/2.5~Z$\sunn\
and $Z = 1/20~Z$\sunn\ $EW$(H$\beta$) decreases to a value of 50~\AA\
(roughly corresponding to $EW$([\ion{O}{iii}]\,$\lambda$\,5007)
$\simeq 200$~\AA, see
Section~\ref{Discussion} and Fig.~\ref{EWs_HbO3}) in 4.7 and
7~Myr, respectively. Suppose that we have an ensemble of BCG progenitors
with a broad metallicity distribution. We suggest that these progenitors
experience instantaneous starbursts with a quickly fading strength of
emission lines. Then, due to the above effect, for the fixed threshold value
of EW(H$\beta$), the lower
metallicity objects will be overrepresented in comparison to their real
fraction among the progenitors. This additionally justifies the search for
the most metal-deficient galaxies (with $Z$ down to 1/50~$Z$\sunn) among
strong-lined BCGs.

In reality the situation is more complicated. The mean ratio of
I([\ion{O}{iii}]\,$\lambda$\,5007~\AA)/I(H$\beta$) (or equivalently,
the ratio of their EWs, since these lines are rather close in wavelength)
is a function of metallicity. It peaks near $Z \sim 1/10~Z$\sunn\ at the
level of $\sim 5-10$, as seen in the compilation and models of Stasinska
\& Leitherer (\cite{SL96}).
For BCGs with $Z \sim (1/20-1/30)~Z$\sunn\ this ratio varies between $\sim 2$
and 6.5 (see, e.g., data in Skillman et al. \cite{Skillman88},  Salzer et al.
\cite{Salzer91}, Izotov et al. \cite{Izotov94}, van Zee et al. \cite{Zee96},
Kniazev et al. \cite{Kniazev98}, Thuan et al. \cite{Thuan99},  Kniazev et al.
\cite{Kniazev00a, Kniazev00b}, Fricke et al. \cite{Fricke01}, and in this
paper).
For galaxies with $Z$ as low as $\sim 1/40-1/50~Z$\sunn\ it falls
to $\sim 1.5-3$ (see data for I~Zw~18, SBS~0335--052~E and W, UGCA~292, e.g.,
in Izotov \& Thuan \cite{Izotov99}, Lipovetsky et al. \cite{Lipovetsky99},
van Zee \cite{Zee00}). For \ion{H}{ii} regions
with metallicities near the solar value this ratio is usually found
to be less than $\sim 2$ (e.g., van Zee et al. \cite{Zee98c}).

Therefore the threshold value of EW([\ion{O}{iii}]\,$\lambda$\,5007)
of 200~\AA\ for the
extremely metal-deficient galaxies will correspond to a threshold in
EW(H$\beta$) of $50-130$~\AA, while for more typical BCGs this threshold in
EW(H$\beta$) will be only $35-40$~\AA. This will compensate, and even
counteract (for the lowest metallicities we are seeking for, namely
$Z \sim 1/50~Z$\sunn), the effect of a longer EW(H$\beta$) decay,
mentioned above.
It is also useful to note that the primary selection criteria, which we
impose for LRS, work against higher metallicity objects. In particular, as
it is seen in the bottom panel of Figure~\ref{Fig2}, we miss about half
of the BCGs with $Z > 1/5~Z$\sunn\ but none of those with $Z < 1/10~Z$\sunn.
All these factors should later be properly accounted for while  
the O/H distribution is analyzed with the aim to obtain an unbiased
metallicity distribution.

Such strong-lined ELGs have a very low surface density. Thus, a sky region
with an area of several thousand square degrees should be examined in order
to end up with a sample of the order of a hundred BCGs.
It is crucial therefore to find the optimal region in the two-parameter space
(slope ({\it slo}), density ({\it dns})) used for automatic selection in the
basic data-set of LRS spectra, to minimize the number of candidates, keeping
at the same time the loss of strong-lined BCGs low.
Based on the training BCG sample taken from the Second Byurakan Survey
(see description of the sample in Ugryumov et al.~\cite{Ugryumov99})
we determine this region as follows (Fig.~\ref{Fig2}):

\begin{enumerate}

\item The LRS continua should be sufficiently blue, e.g.
      $slo \ge -0.1$. \\

\item The integrated density (roughly corresponding to the
      object brightness) should be sufficently high, in order to minimize
      spurious ``emission lines'' due to noise peaks at low S/N, e.g.
      $dns \ge 150$.

\end{enumerate}

The more stringent restriction of  the ``slope'' parameter compared to the
HSS $slo \ge -0.3$ allows us to avoid the region populated with lower
excitation ELGs. This in turn leads to a large reduction in the number of
spectra surviving this step of the selection procedure:
the number of candidates decreases from $10\,000-12\,000$ per plate in the
HSS down to $2500-3500$ per plate in the HSS--LM.

In contrast to the HSS an additional selection step using LRS data is
therefore avoided.
However, some strong-lined BCGs will then be missed and, after analysing
the LRS parameters of $\sim 100$ strong-lined BCGs from the HSS and the SBS,
we estimate this loss to be of the order 10~\%.
The next step of the selection procedure is performed by visual inspection
of the high-resolution scans (HRS) of the candidates on a graphic display.
Only candidates with really strong lines were selected.

Our experience from the HSS showed that galaxies with
EW([\ion{O}{iii}]\,$\lambda$\,5007) $>$~$150-200$~\AA\ always show a strong
signature in the high-resolution scans, making the interactive selection
a non-critical step with respect to bias.

One of the advantages of the selection work performed for the
HSS--LM is that for each field one HQS prism plate already has been fully
digitized with high resolution and is stored in a compact disk database.
Therefore the time usually spent for rescanning individual spectra
with high resolution was saved. On the other hand, there were some
disadvantages related to the fact that only one (the best quality) prism
plate per field was digitized with high resolution.
This leads to a small number of emission-line candidates in which the
emission features are in fact artifacts caused either by dust grains or by
possible ghosts or noise peaks. All such cases were detected in the course
of applying additional selection procedures, as described below.

\subsection{Additional selection with the use of the APM database and
DSS-II images}
\label{APM}

\begin{figure}
\centering
\vspace{-0.6cm}
\psfig{figure=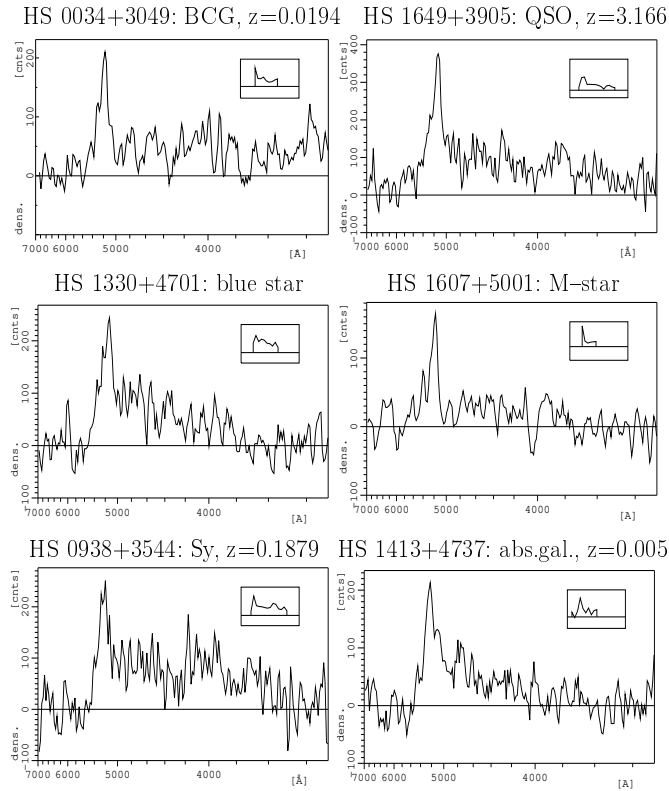,width=9.6cm,angle=0}
\vspace{-2cm}
\caption[]{ \label{HRS_exampl}
 Examples of faint HRS spectra selected as candidate
 strong-lined ELGs and their classifcation after follow-up spectroscopy:
 BCG, QSO, blue star, M-star, AGN and absorption-line galaxy.
}
\vspace{-0.4cm}
\end{figure}

The fainter candidates ($B \gtrsim 18\fm0$) picked up from the automatic
selection procedure and visual inspection still contain many objects
of unwanted types. In Fig.~\ref{HRS_exampl} we show how similar the HRS
spectra of very different types of objects (BCG, QSO, blue star, M-star,
AGN and absorption-line galaxy) can look.

Using the APM database (Irvin \cite{Irwin98}) to separate star-like and
extended objects as well as to get the APM~$(B-R)$ colour is quite helpful
to reject further candidates. According to our previous experience with
the selection of the HSS candidates,
the great majority of APM star-like candidates with a ``strong emission''
feature in the HRS near $\lambda$\,5000\,\AA\ appeared to be either
M-stars (in case of ``red'' APM colours $(B-R) > 2$), or quasars with
$z\approx 3$ (in case of ``neutral'' APM colours $(B-R) \sim 1.0-1.7$)
or B,A-stars with sufficiently strong H$\beta$ absorption
(in case of ``blue'' APM colours $(B-R) < 1$)).

``Blue'' star-like objects have been additionally checked on the images of
the Digitized Sky Survey (DSS-II) because some very compact BCGs could be
hidden among them. All doubtful cases were kept in the list for follow-up
spectroscopy with the 6\,m telescope. In order to save observing time
these objects were first observed with a short exposure ($\sim 1-3$~min)
and only after confirmation of their true ELG nature a ``normal'' exposure
($\sim 20$~min) was made.

One more category of non-BCG non-stellar strong-lined candidates was
recognized with the help of the APM database and was removed from the
observational list. These are APM ``red'' galaxies ($(B-R) > 1.7$)), which
from our earlier experience from the work with the HSS, and from
cross-checking of our candidates with available spectra from the Sloan
Digital Sky Survey Early Release Data (EDR), turned out to be  either
absorption-line galaxies with $z \sim 0.1$, or AGN type
galaxies. A small number of strong-lined BCGs can occasionally appear
in this range of $(B-R)$. From the statistics of HSS BCGs we estimate that
we miss $< 3$~\% of such objects.

\section{Spectral observations and data reduction}

\subsection{The 6\,m telescope spectroscopy}

The results presented here were obtained from the observations with
the Russian 6\,m telescope during 6 runs between July 1999 and October 2000.
The Long-Slit Spectrograph (LSS in Table~\ref{Tab1}; Afanasiev et al.
\cite{Afanasiev95}) attached to the telescope prime focus was used.
Most of the long-slit spectra
(1\farcs2~--~2\farcs0$\times$120\arcsec) were obtained with the grating of
651 grooves~mm$^{-1}$, giving a dispersion of 2.4~\AA\,pixel$^{-1}$.
The data from the first and fourth runs were obtained with the grating of
325 grooves~mm$^{-1}$ giving a dispersion of
4.6~\AA\,pixel$^{-1}$. The scale along the slit in both set-ups was
0\farcs39~pixel$^{-1}$.
For all observations the Photometrics CCD-detector PM1024 was used with a
$24\times24\,\mu$m pixel size.

Normally, one or two exposures per object ($10-20$~minutes for one exposure
depending on brightness and observational conditions) were used
in order to detect the
[\ion{O}{iii}]\,$\lambda$\,4363~\AA\ emission line and to measure the oxygen
abundance by the  classical standard method. For the most
metal-deficient galaxies we got additional spectroscopy with a higher S/N
ratio, to confirm their very low metallicity. For about ten doubtful star-like
candidates we got $1-3$~min. exposure spectra to be sure we did not miss
some very compact BCGs. For strong-lined galaxies the three strongest lines
are well seen in such spectra.
Reference spectra of an He--Ne--Ar lamp were recorded
before or after each observation to provide the wavelength calibration.
Spectrophotometric standard stars from Oke (\cite{Oke90}) and Bohlin
(\cite{Bohlin96}) were observed for the flux calibration at least twice a
night. All observations and data acquisition have been conducted using the
{\tt NICE} software package by Kniazev \& Shergin (\cite{Kniazev95}) in
the MIDAS\footnote{MIDAS is an acronym for the European Southern
Observatory package --- Munich Image Data Analysis System.} environment.

All observations with exception of two nights were conducted during good
photometric conditions with a seeing of $1\farcs0-2\farcs0$. The two nights
on August 12, 1999 and on July 29, 2000 (during which the spectra of
5 objects have been obtained) were non-photometric, with the seeing of
$3\farcs0-4\farcs0$. For two of them spectra of higher quality were acquired
in following runs.

\subsection{Data reduction}

The reduction of the spectra was performed at the SAO using the standard
reduction systems MIDAS and IRAF\footnote[2]{IRAF is distributed by
National Optical Astronomical Observatories, which is operated by the
Association of Universities for Research in Astronomy, Inc., under
cooperative agreement with the National Science Foundation}.

Two-dimensional CCD images were automatically cleared from
cosmic ray hits with the MIDAS routine FILTER/COSMIC.
We then applied the IRAF package CCDRED to perform bad pixel removal,
trimming, bias-dark subtraction, slit profile and flat-field corrections.

Prepared images were processed with the IRAF package LONGSLIT
for wavelength calibration, distortion and tilt correction of each frame
followed by sky substraction and correction for atmospheric extinction.
This standard way of reduction was performed by the routines: IDENTIFY,
REIDENTIFY, FITCOORD, TRANSFORM, BACKGROUND and EXTINCTION.

To perform the flux calibration for all object images, the instrumental
response curves were obtained after the reduction of the spectrophotometric
standard star spectra.
This reduction includes as a first step extraction of apertures of standard
stars with the use of the APSUM procedure from the APEXTRACT package.
In a second step STANDARD and SENSFUNC procedures transformed these
apertures into sensivity curves for the CALIBRATE procedure to perform
flux calibration for all two-dimensional object images.
To extract one-dimensional spectra from the flux calibrated images
the APSUM routine was used.
In the case that more than one exposure was obtained with the same setup
for an object, the extracted spectra were averaged.

For speeding-up and facilitating line measurements we employed
command files created at the SAO using the FIT context and MIDAS
command language. Their description was given in detail in
Papers III-IV of the HSS.

\begin{figure}
\centering
\psfig{figure=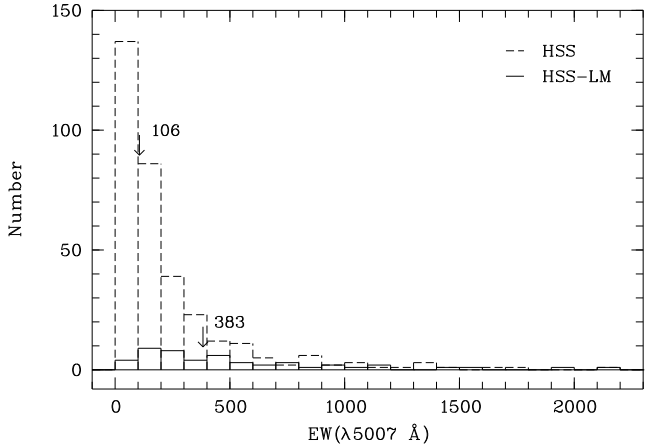,width=9cm,angle=-90}
\vspace{-0.2cm}
\caption[]{\label{Fig3}
 The distribution of $EW$([\ion{O}{iii}]\,$\lambda$\,5007) for all BCGs from
 this work (solid line) and similar data for all BCGs from the HSS
 (Paper~I--V) (short dashed line). The arrows indicate median values.
 The number of ELG per bin in the HSS--LM increases, as expected,  with the
 decrease of $EW$, reaching the maximum in the bin of $100-200$~\AA\
 indicating that the completeness limit is in this range.
 For the HSS BCGs the
 number per bin grows till the last bin ($0-100$~\AA).
 Since for the strong-lined BCGs/\ion{H}{ii}--galaxies a rather tight
 correlation exists between $EW$([\ion{O}{iii}]\,$\lambda$\,5007)
 and $EW$(H$\beta$) (see Fig.~\ref{EWs_HbO3}),
 we well pick up galaxies with $EW$(H$\beta$) $>$~50~\AA. The
 latter limit corresponds to ages of instantaneous SF bursts of
 $T_\mathrm{burst} < 4.7-7$~Myr for the metallicity
 range of 1/20 to 1/2.5~$Z$\sunn\ (Schaerer \& Vacca \cite{SV98}).
}
\end{figure}

\subsection{Oxygen abundances}

The line fluxes were measured applying Gaussian fitting and
integrating over the fitted profile with the INTEGRATE/LINE command
under MIDAS. For blended lines, such as H$\alpha$,
[\ion{N}{ii}]\,$\lambda\lambda$\,6548,6583~\AA,
[\ion{S}{ii}]\,$\lambda\lambda$\,6716,6731~\AA\ and
H$\gamma$, [\ion{O}{iii}]\,$\lambda$\,4363~\AA\ a deblending procedure was
used, assuming Gaussian profiles with the same FWHM as for single lines.
The errors of the line intensities take into account
the Poisson noise statistics and the noise statistics in the continuum
near each line, and include uncertainties of data reduction.
These errors have been propagated to calculate element abundances.
For the simultaneous derivation of $C$(H$\beta$) and $EW$(absorption),
and to correct for extinction, we used the procedure described in detail
by Izotov, Thuan \& Lipovetsky (\cite{Izotov94}).
The abundances of the ionized species and the total abundances
of oxygen and other elements have been obtained
following Izotov et al.~(\cite{Izotov94}, \cite{Izotov97})
and Izotov \& Thuan (\cite{Izotov99}), which in turn are based
on formulae from Aller (\cite{Aller84}), and account for 
$T_\mathrm{e}$ differences in the zones with the predominant O$^{++}$ and
O$^{+}$. In more detail the whole procedure is described in
papers of Kniazev et al.~(\cite{Kniazev00b}) and Pustilnik et al.~
(\cite{PKM02}).

\section{Results}
\label{Results}

We present here the data of 51 emission-line galaxies and one quasar
which were observed in the frame of the HSS--LM in $1999-2000$.
16 of these ELGs were already listed in the NED either as galaxies or as
objects from various catalogs.
For 12 of them NED presents radial velocities.
For a few of these twelve galaxies there is also information in
earlier publications on the presence of emission lines in their spectra,
but there were no data on O/H except UGCA20.
All these objects were included into our observing program as
independently selected strong-lined candidates, expected to have a detectable
[\ion{O}{iii}]\,$\lambda$\,4363~\AA\ line, and thus suitable for O/H determination.
Comparison of our newly determined velocities (Table 2 and 3) with those for
galaxies with already known redshifts shows a good consistency within the
uncertainties given.

\subsection{Emission-line galaxies}

The results of the reduction and the galaxy spectra analysis are summarized in
Tables~\ref{Tab2} and \ref{Tab3}. We give only integrated parameters of
the galaxies studied, including the value of the oxygen abundance.
For 24 of the galaxies presented  various indications for characteristic
WR features are detected. Of them 8 show a definite blue WR bump near
$\lambda\,4600-4700$~\AA\ only, another 7  show both blue and red
($\lambda\,5800$~\AA) WR bumps. 9 further ELGs show less confident WR blue
bumps. This is consistent with the young ages of the SF bursts implied by
their large values of $EW$ of H$\beta$. The results of a more detailed
analysis, including all observed line intensities and other heavy element
abundances, as well as detailed information on the detected Wolf-Rayet
features will be presented in forthcoming papers.

The strong-lined ELGs are listed in Table~\ref{Tab2}, while the ELGs with
fainter lines
($EW$([\ion{O}{iii}]\,$\lambda$\,5007) $<$ 100~\AA) are separated in
Table~\ref{Tab3}.
Tables~\ref{Tab2} and \ref{Tab3} contain the following information: \\
 {\it column 1:} The object's IAU-type name with the prefix HS. \\
 {\it column 2:} Right ascension for equinox J2000. \\
 {\it column 3:} Declination for equinox J2000.
The coordinates were measured on direct plates of the HQS
and are accurate to $\sim 2\arcsec$ (Hagen et al. \cite{Hagen95}). \\
 {\it column 4:} Heliocentric velocity in \kms. \\
 {\it column 5:} rms of the heliocentric velocity uncertainty in \kms. \\
 {\it column 6:}
Blue appparent magnitudes based on the values from the APM database.
The latter were corrected following Kniazev et al. (\cite{Kniazev02}) to
transfer them to the Johnson $B$-band system. Their rms uncertainty
is estimated as 0\fm45. \\
 {\it column 7:} The extinction $A_{B}$ in our Galaxy in the $B$-band,
(Schlegel et al. 1998). \\
 {\it column 8:} The absolute $B$-magnitudes are calculated from the apparent
ones in {\it column 6} and the heliocentric velocities and are corrected
for $A_{B}$ in {\it column 7}. \\
 {\it columns 9,10:} The equivalent widths of the emission lines
H$\beta$ and [\ion{O}{iii}]\,$\lambda$\,5007 in \AA ngstr\"om. \\
 {\it column 11} (Table~\ref{Tab2}): The derived values of
12+$\log$(O/H). Their rms uncertainties vary from 0.02 dex for the
largest S/N of the [\ion{O}{iii}]\,$\lambda$\,4363~\AA\ line to $\sim 0.2$ dex for
the lowest S/N (see plots of the spectra in Appendix~A).
The most uncertain data ($\sim 0.2$ dex) are marked by a colon. \\
 {\it column 12} (Table~\ref{Tab2}) and {\it column 11} (Table~\ref{Tab3}):
One or more alternative names, according to information from NED.
Reference numbers are given to other sources for
redshift-spectral information indicating that a galaxy is an ELG.

According to our spectral classification all, but two of the observed
ELGs turned out to be BCGs with a characteristic \ion{H}{ii}-region spectrum
and low luminosity. The only exceptions are the two galaxies HS~2300+2612
and HS~0053+2910. The former is classified as Starburst Nucleus (SBN),
the latter -- as Dwarf Amorphous Nucleus Starburst (DANS).
The classification criteria applied follow those of Salzer~(\cite{Salzer89a}).
Here Salzer's classes SS, DHIIH and HIIH are taken together as one
class of blue compact/HII galaxies (BCG) (see also Ugryumov et al.
~\cite{Ugryumov99}).

The spectra of all emission-line galaxies are shown in Appendix~A,
which is available only in the electronic version of the journal.

Since for part of the observed galaxies the
[\ion{O}{iii}]\,$\lambda$\,4363~\AA\ line was only barely detected and the
derived O/H value is of low precision, we have
checked these objects with the empirical strong-line method devised by
Pilyugin (\cite{Pilyugin00,Pilyugin01}), giving an independent estimate.
In all cases but one we got consistent results within the cited uncertainties.
The one exclusion is HS~2259+2357, for which we have an estimate of
12+$\log$(O/H)=$8.0\pm0.2$, while Pilyugin's method gives 8.6.

\subsection{UGCA~20}

This galaxy was selected under the name HS~0140+1943 as a candidate
strong-lined object. In the process of
cross-check it was identified in NED as the galaxy UGCA~20, for which
high S/N spectroscopy was obtained by van Zee et al. (\cite{Zee96}).
From the high S/N spectra of its two identified \ion{H}{ii} regions they
determined an oxygen abundance of 12+$\log$(O/H)=$7.58\pm0.03$.
Our detailed study of this galaxy (a paper in preparation) resulted in
an even lower metallicity: 12+$\log$(O/H)=$7.42\pm0.06$.

\subsection{Quasar HS~0057+2049}

This new faint ($m_B$=18\fm7) quasar was observed as a ``blue'' star-like
candidate (see the end of Section~\ref{sample}) to a possible very compact
BCG. Its coordinates are  $\alpha_{2000}$=$01^h 00^m 27^s.7$,
$\delta_{2000}$=$+21^\circ 05' 42''$. The strong and broad emission line at
$\lambda$~$\approx$ 4980~\AA\ is unambiguously identified with
Ly$\alpha$, what gives a redshift of $3.0983\pm0.0004$.
The finding chart of this quasar and the plot of its spectrum can be found
on the www-site of the Hamburg Quasar Survey
(http://www.hs.uni-hamburg.de/hqs.html).

\begin{figure}
\centering
\psfig{figure=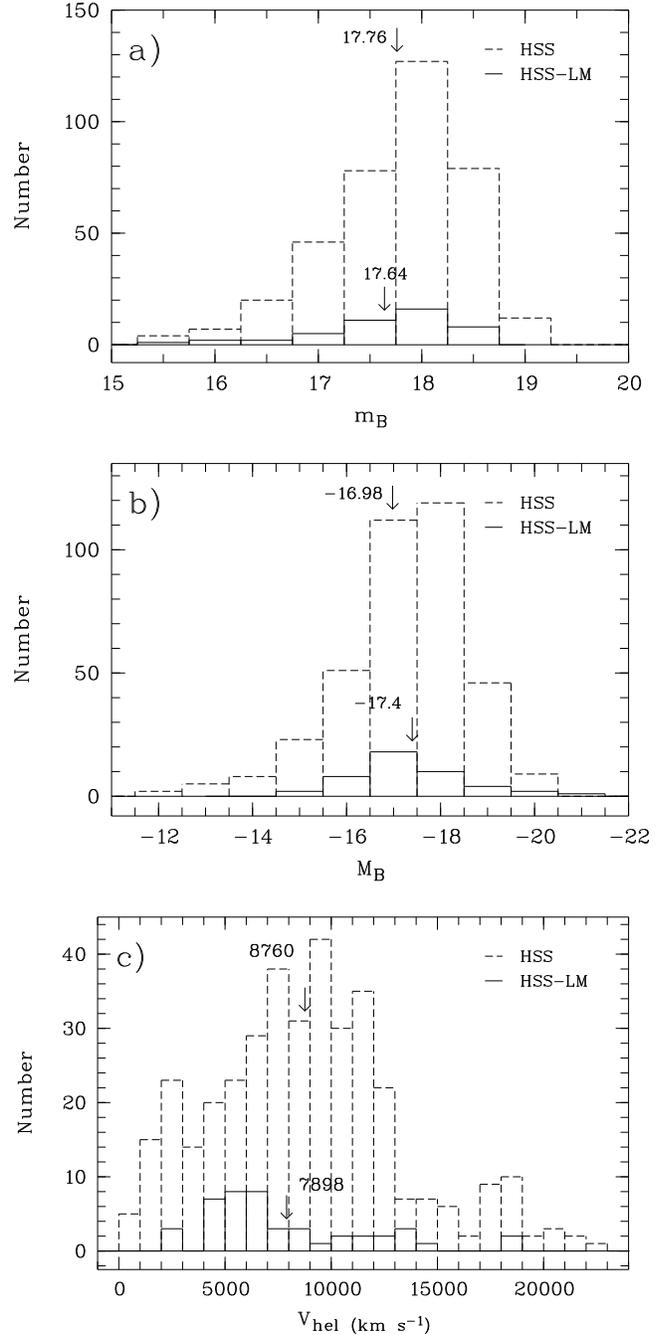,width=13.3cm,angle=0}
\vspace{-0.5cm}
\caption[]{\label{Fig4}
 Distributions of apparent ({\bf a}), absolute ({\bf b}) $B$-magnitudes
 and of $V_\mathrm{hel}$ ({\bf c}) for the sample of HSS--LM BCGs with
 $EW$([\ion{O}{iii}]\,$\lambda$\,5007) $\ge$~100~\AA\ in comparison to
 all BCGs from the HSS (Paper~I--V). The arrows indicate mean values.
 The Figures show the trend of these low-mass galaxies to be well sampled
 up to $m_B\sim 18$ and $V_\mathrm{hel} \sim 7000$~\kms.
}
\end{figure}

\begin{figure}
\centering
\psfig{figure=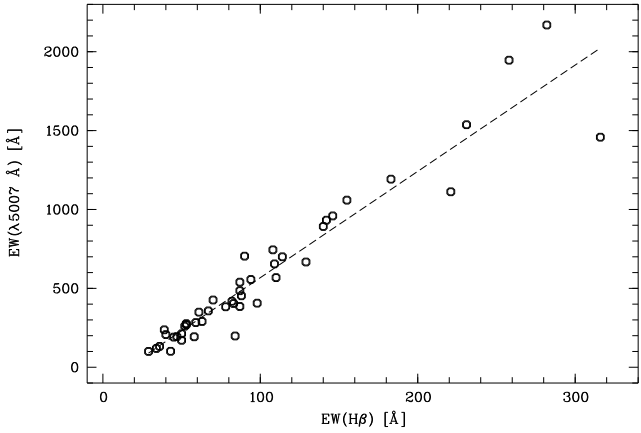,width=9cm,angle=-90}
\caption[]{\label{EWs_HbO3} The observed relation between the
EW([\ion{O}{iii}]\,$\lambda$\,5007) and EW(H$\beta$) of 46 strong-lined BCGs
from Table~\ref{Tab2}. The dashed line is a linear fit to the data.
The threshold value of EW([\ion{O}{iii}]\,$\lambda$\,5007)~=~200~\AA\
for the expected completeness limit of the sample corresponds to
EW(H$\beta$)~=~50~\AA.
}
\end{figure}

\begin{figure}
\centering
\psfig{figure=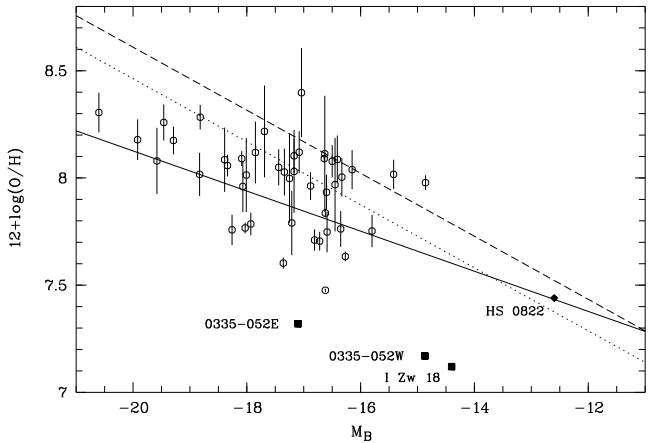,width=9cm,angle=-90}
\caption[]{\label{Z_Lum} The strong-lined BCGs (open circles) from
Table~\ref{Tab2} plotted
in the well known diagram metallicity vs blue luminosity (e.g. Skillman et al.
\cite{Skillman89}), originally shown for dwarf irregular galaxies.
The dashed line shows the original relation found by Skillman et al.
(\cite{Skillman89}), the dotted line shows that relation shifted to the left
by 1$^\mathrm{m}$ to account for the expected brightening of BCGs. The solid
line is a linear fit to our observed data with the weights accounting for O/H
uncertainties. The error bars of O/H correspond to $\pm 1\sigma$.
Pay attention that three of the most metal-deficient BCGs (not taken for
the estimate of this linear fit), shown by filled squares, strongly deviate
from any of the fitting lines.
}
\end{figure}

\section{Discussion and conclusions}
\label{Discussion}

\subsection{Current results and relation to the goals}

We have presented above the description of the new project and its first
results. The extremely low-metallicity galaxies are very rare objects, and
the discovery of a few such objects, probably does not well justify the
efforts undertaken. Therefore, another important goal of the project is to
create a large BCG sample with reliably measured O/H and a well-defined
selection function.

The first goal of the project is shown to be reached given the
discovered galaxies with O/H lower than 1/20 of the solar value.
Three such very rare objects were discovered, and one more was rediscovered.
We expect that among the remaining candidates a comparable number of
extremely metal-poor galaxies will be found. Thus, HSS--LM
will substantially increase the whole number of these unusual objects, and
will advance the study of their population properties and the search for
truly young local dwarf galaxies.

As for the second goal the results presented above allow us to
formulate some preliminary conclusions and to estimate its perspectives.

First, the selection procedures elaborated, based on the Hamburg Quasar
Survey LRS and HRS databases, supplemented by information
from the APM database and DSS II images, yields a very high fraction of
strong-lined BCGs. Namely, 37 ELGs with
$EW$([\ion{O}{iii}]\,$\lambda$\,5007) $\ge$ 200~\AA\ out of 52 observed
candidates (that is 71~\%) have been found (see also Fig.~\ref{Fig3}).
This confirms the high efficiency of the applied selection method.

Second, a preliminary analysis of the parameters of the new strong-lined BCGs
(see Fig.~\ref{Fig4}),
indicates reasonable completeness of the sample with well determined O/H
in apparent $B$-magnitude up to $\sim 18^\mathrm{m}$.
The distribution of the sample BCGs in radial velocity is increasing
up to $V_\mathrm{hel} = 7000$~\kms. This implies that all BCGs with
$M_{B} \le -17\fm0$ will be picked up with high efficiency. Based on
the HSS experience, the well sampled velocity range is probably larger.

Third, the temperature-sensitive weak line
[\ion{O}{iii}]\,$\lambda$\,4363~\AA\ was
detected in 43 BCGs in a sky region covering $\sim 1/3$ of the entire
HSS--LM area. The oxygen abundances 12+$\log$(O/H) derived for these
BCGs vary in the wide range between $\sim 7.4$ and 8.4, having a typical
uncertainty of $\sim 0.1$~dex.
The number of homogeneously selected BCGs with reliable O/H
value expected in the whole sky zone of the HSS--LM is
$\ga$ $100-120$ galaxies. They will provide a good opportunity to make
first steps toward the understanding a selection-free metallicity
distribution of the local gas-rich galaxies --- BCG progenitors.

\subsection{New sample and its relation to BCGs from the HSS}

The comparison of parameters of the observed BCG subsample from
the HSS--LM with those for BCGs from the HSS (Fig.~5) shows that the
latter are distributed rather similar to the former in apparent blue
magnitude $B$, in absolute $M_B$ and in radial velocity. Their mean
values are quite close: $<B>$ = 17\fm64 and 17\fm76, $<M_B>$ =
$-17\fm40$ and $-16\fm98$, and $<V_\mathrm{hel}>$ = 7898 and 8760~\kms\ for
the HSS--LM and the HSS respectively. Therefore we expect that the well
selected sample of strong-lined BCGs from HSS--LM can be combined with
the subsample of about one hundred strong-lined BCGs in the HSS zone to
derive a metallicity distribution.

\subsection{Preliminary results on parameter correlations}

To illustrate statistical relations between BCG parameters discussed
in this paper and in the literature we show Figures 6 and 7 based on
the data obtained here in this paper.

In Fig.~\ref{EWs_HbO3} we present a
rather tight relation between the $EW$s of H$_{\beta}$ and [\ion{O}{iii}]
$\lambda$\,5007~\AA. This correlation allows to fix the ages of instantaneous
SF bursts, which are well picked up by the applied selection criterion based
on the $EW$([\ion{O}{iii}]$\,\lambda$\,5007) (see Introduction).

The other important relation is often discussed for dIrr galaxies: there is
a general trend of metallicity decrease with decrease of blue luminosity
(e.g. Skillman et al. \cite{Skillman89}, Richer \& McCall \cite{RM95}).
For BCGs, if they are related to
other gas-rich low-mass galaxies, one could expect a similar relation.
Current data, presented in Fig.~\ref{Z_Lum}, show a rather large scatter,
so it is difficult to draw definite conclusions from the strong-lined BCG
data available. The overall trend seen in the data fits well with previous
results (c.f. Kunth \& \"Ostlin (\cite{KO}); their Fig.~10).
It is worth noting however, that the extremely
metal-deficient BCGs strongly deviate from any relation valid for
the rest of the BCGs. This implies that the search for the most metal-poor
BCGs should not be concentrated on the least luminous candidates only.

Summarizing the presented results, we draw the following conclusions:

\begin{enumerate}

\item The HSS--LM is a new HQS based project, which very efficiently samples
      the strong-lined ELGs of \ion{H}{ii}/BCG type.
      After the 6\,m telescope follow-up spectroscopy 71~\% of the
      candidates (37 galaxies) were found to be BCG/\ion{H}{ii}-galaxies with
      $EW$([\ion{O}{iii}]\,$\lambda$\,5007) in the range of 200 to 2100~\AA.

\item Among these galaxies 3 new very metal-deficient BCGs were found
      with $Z$ in the range 1/28 to 1/20~$Z$\sunn. One more such galaxy --
      UGCA~20 -- with $Z =$ 1/32~$Z$\sunn\ was rediscovered as one
      of the HSS--LM targets.

\item For 43 strong-lined BCGs the oxygen abundance was determined with
      the use of the temperature sensitive faint line
      [\ion{O}{iii}]\,$\lambda$~4363~\AA.
      Its full range corresponds to ionized gas metallicities between
      1/28 and 1/3~$Z$\sunn.

\end{enumerate}

This paper presents about 1/3 of the galaxies from the final sample of
strong-lined BCGs of this survey. The observations of the remaining
objects are expected to be completed during the years $2002-2003$.

\begin{acknowledgements}

A.V.U. is very grateful to the staff of the Hamburg Observatory for
their hospitality and kind assistance.  We acknowledge the partial
support from INTAS (grant 97-0033), Russian state program
``Astronomy'' and Center of Cosmoparticle Physics ``Cosmion''.
The authors are thankful to the referee G.\"Ostlin for useful suggestions.
This research has made use of the NASA/IPAC Extragalactic Database
(NED) which is operated by the Jet Propulsion Laboratory, California
Institute of Technology, under contract with the National Aeronautics
and Space Administration. The use of the Digitized Sky Survey (DSS-II)
and APM Database is gratefully acknowledged.

\end{acknowledgements}

\renewcommand{\baselinestretch}{0.5}

\clearpage

\begin{landscape}
\headsep 7cm
\renewcommand{\baselinestretch}{0.8}


\setcounter{qub}{0}

\begin{table*}[h]
                                                               
\begin{center}

\caption{\label{Tab2} Parameters of the strong-lined BCGs}

\vspace{-0.3cm}

\begin{tabular}{rlccrlcccrrcl} \hline \\[-0.2cm]
\multicolumn{1}{c}{\#}               &                         
\multicolumn{1}{c}{Name}             &                         
\multicolumn{1}{c}{$\alpha\,(2000)$} &
\multicolumn{1}{c}{$\delta\,(2000)$} &
\multicolumn{1}{c}{V$_\mathrm{hel}^{\;\;\;\;a}$}  &
\multicolumn{1}{c}{$\pm\sigma_\mathrm{V}$}  &
\multicolumn{1}{c}{m$_{B}^{\;\;b}$}  &                         
\multicolumn{1}{c}{$A_{B}^{\;\;c}$}  &
\multicolumn{1}{c}{M$_{B}^{\;\;d}$}  &
\multicolumn{1}{r}{EW$_\mathrm{H\beta}$}       &
\multicolumn{1}{r}{EW$_{\lambda 5007}$} &
\multicolumn{1}{c}{12+$\log$(O/H)$^{\,e}$} &
\multicolumn{1}{l}{Other names from NED} \\
&
\multicolumn{1}{c}{ }  &
\multicolumn{1}{c}{ }  &
\multicolumn{1}{c}{ }  &
\multicolumn{1}{c}{(km s$^{-1}$)}  &
\multicolumn{1}{c}{(km s$^{-1}$)}  &
\multicolumn{1}{c}{ }  &
\multicolumn{1}{r}{ }  &
\multicolumn{1}{r}{ }  &
\multicolumn{1}{r}{(\AA) }  &
\multicolumn{1}{r}{(\AA) } &
\multicolumn{1}{c}{ } &
\multicolumn{1}{l}{and number of reference} \\
&
\multicolumn{1}{c}{ (1) }  &                                   
\multicolumn{1}{c}{ (2) }  &                                   
\multicolumn{1}{c}{ (3) }  &                                   
\multicolumn{1}{c}{ (4) }  &                                   
\multicolumn{1}{c}{ (5) }  &
\multicolumn{1}{c}{ (6) }  &
\multicolumn{1}{r}{ (7) }  &
\multicolumn{1}{r}{ (8) }  &
\multicolumn{1}{r}{ (9) }  &
\multicolumn{1}{c}{ (10)}  &
\multicolumn{1}{c}{ (11)}  &
\multicolumn{1}{c}{ (12)} \\
\\[-0.2cm] \hline \\[-0.2cm]
\qq& HS 0017+1055   & 00 20 21.4 & +11 12 21 &  5650 & $\pm$ 12 & 18.26 & 0.15 & --16.27 & 146 & 959  & 7.63 & [SS98] 84, {\bf 1} \\
\qq& HS 0024+2314   & 00 26 52.4 & +23 31 13 &  6967 & $\pm$ 10 & 17.51 & 0.10 & --17.44 & 40  & 207  & 8.04 & \\
\qq& HS 0029+1443   & 00 32 18.5 & +15 00 11 &  5325 & $\pm$ 17 & 17.52 & 0.15 & --16.88 & 87  & 485  & 7.96 & {\bf 1} \\
\qq& HS 0029+1748   & 00 32 03.2 & +18 04 44 &  2188 & $\pm$ 30 & 17.62 & 0.15 & --14.86 & 183 & 1192 & 7.97 & NPM1G +17.0024, {\bf 1} \\
\qq& HS 0031+2645   & 00 34 19.7 & +27 02 09 &  4105 & $\pm$ 12 & 18.07 & 0.15 & --15.80 & 53  & 270  & 7.75 & \\
\qq& HS 0034+3047   & 00 37 36.8 & +31 04 08 &  5825 & $\pm$ 45 & 18.22 & 0.19 & --16.45 & 50  & 212  & 7.96:& \\
\qq& HS 0041+2333   & 00 44 20.5 & +23 50 00 &  6554 & $\pm$ 24 & 16.76 & 0.11 & --18.07 & 61  & 349  & 7.96 & [SS98] 88, {\bf 1} \\
\qq& HS 0044+3052   & 00 47 26.9 & +31 09 02 &  5209 & $\pm$ 13 & 17.77 & 0.16 & --16.63 & 50  & 171  & 8.11:& \\
\qq& HS 0048+2744   & 00 51 05.9 & +28 00 35 &  6974 & $\pm$ 10 & 17.71 & 0.19 & --17.34 & 70  & 426  & 8.02 & \\
\qq& HS 0051+2812   & 00 54 23.5 & +28 29 09 & 18490 & $\pm$ 12 & 17.83 & 0.15 & --19.29 & 78  & 383  & 8.17 & \\
\qq& HS 0052+2536   & 00 54 56.4 & +25 53 08 & 13606 & $\pm$ 15 & 16.85 & 0.13 & --19.58 & 34  & 119  & 8.07:& IRAS F00522+2537 \\
\qq& HS 0052+2537   & 00 54 56.0 & +25 53 23 & 13737 & $\pm$ 12 & 18.44 & 0.13 & --18.01 & 82  & 418  & 8.01 & IRAS F00522+2537 \\
\qq& HS 0058+1847   & 01 01 32.3 & +19 03 33 & 11192 & $\pm$ 19 & 18.10 & 0.08 & --17.85 & 59  & 283  & 8.11 & {\bf 1} \\
\qq& HS 0103+3219   & 01 06 03.6 & +32 35 32 &  5341 & $\pm$ 10 & 18.31 & 0.17 & --16.15 & 83  & 404  & 8.03 & \\
\qq& HS 0109+3304   & 01 12 30.6 & +33 20 04 &  4846 & $\pm$ 10 & 17.74 & 0.16 & --16.50 & 67  & 357  & 8.07 & KUG 0109+330 \\
\qq& HS 0111+2115   & 01 14 37.6 & +21 31 16 &  9445 & $\pm$ 12 & 16.14 & 0.10 & --19.46 & 110 & 568  & 8.25 & NPM1G +21.0056, {\bf 1} \\
\qq& HS 0113+1750   & 01 16 40.3 & +18 05 51 & 18756 & $\pm$ 21 & 18.25 & 0.08 & --18.82 & 155 & 1059 & 8.28 & {\bf 1} \\
\qq& HS 0119+3059   & 01 22 14.6 & +31 15 16 &  4793 & $\pm$ 27 & 17.64 & 0.19 & --16.60 & 87  & 539  & 7.93 & \\
\qq& HS 0121+1753   & 01 24 00.0 & +18 09 04 &  7760 & $\pm$ 12 & 18.01 & 0.11 & --17.17 & 52  & 260  & 8.10 & \\
\qq& HS 0122+0743   & 01 25 34.2 & +07 59 22 &  2922 & $\pm$ 10 & 15.68 & 0.14 & --17.36 & 221 & 1112 & 7.60 & UGC 00993, {\bf 2} \\
\qq& HS 0123+1624   & 01 26 17.3 & +16 40 29 &  8720 & $\pm$ 10 & 17.12 & 0.14 & --18.34 & 94  & 556  & 8.05 & {\bf 1} \\
\qq& HS 0127+2706   & 01 29 47.0 & +27 22 20 & 12566 & $\pm$ 12 & 15.75 & 0.23 & --20.60 & 98  & 406  & 8.30 & \\
\qq& HS 0128+2832   & 01 31 21.3 & +28 48 12 &  4833 & $\pm$ 36 & 17.62 & 0.20 & --16.64 & 258 & 1946 & 8.09 & \\
\qq& HS 0131+1649   & 01 33 59.1 & +17 04 44 & 11124 & $\pm$ 12 & 17.17 & 0.15 & --18.83 & 63  & 290  & 8.01 & \\
\qq& HS 0134+3415   & 01 37 13.8 & +34 31 12 &  5800 & $\pm$ 30 & 17.95 & 0.11 & --16.62 & 282 & 2169 & 7.83 & \\
\qq& HS 0137+2935   & 01 40 24.1 & +29 51 05 &  4669 & $\pm$ 48 & 17.33 & 0.15 & --16.81 & 129 & 667  & 7.71 & \\
\qq& HS 0137+3152   & 01 40 41.8 & +32 07 19 &  8039 & $\pm$ 12 & 18.12 & 0.17 & --17.21 & 29  & 100  & 7.79:& \\  
\qq& HS 0140+1943   & 01 40 30.3 & +19 43 47 &   574 & $\pm$ 20 & 15.78 & 0.06 & --13.56 & 73  & 246  & 7.42 & UGCA 20, {\bf 3} \\
\qq& HS 0143+2400   & 01 45 54.4 & +24 15 55 & 10350 & $\pm$ 19 & 18.00 & 0.33 & --18.03 & 140 & 892  & 7.76 & {\bf 1} \\
\qq& HS 0205+3212   & 02 08 28.1 & +32 27 05 &  5044 & $\pm$ 15 & 18.10 & 0.28 & --16.33 & 90  & 704  & 8.00 & \\
\qq& HS 0248+2001   & 02 51 06.8 & +20 13 42 &  6402 & $\pm$ 18 & 17.95 & 0.39 & --17.08 & 45  & 191  & 8.12 & \\
\qq& HS 2236+1344   & 22 38 31.1 & +14 00 29 &  6183 & $\pm$ 12 & 18.15 & 0.17 & --16.62 & 316 & 1458 & 7.47 & \\
\qq& HS 2252+1752   & 22 55 15.3 & +18 08 35 & 12747 & $\pm$ 33 & 16.48 & 0.24 & --19.92 & 87  & 385  & 8.17 & \\
\qq& HS 2252+2032   & 22 55 11.6 & +20 48 48 & 13764 & $\pm$ 12 & 18.58 & 0.18 & --17.93 & 142 & 932  & 7.78 & \\
\qq& HS 2256+2610   & 22 58 57.2 & +26 26 23 &  7833 & $\pm$ 39 & 17.61 & 0.18 & --17.69 & 84  & 198  & 8.21:& IRAS F22565+2610, \\
   &                &            &           &       &          &       &      &         &     &      &      & 2MASXi J2258572+262623 \\
\qq& HS 2258+2046   & 23 00 32.2 & +21 02 30 &  5588 & $\pm$ 19 & 18.02 & 0.22 & --16.59 & 88  & 453  & 7.74 & \\
\qq& HS 2258+2215   & 23 00 43.2 & +22 31 13 & 10629 & $\pm$ 39 & 17.55 & 0.17 & --18.39 & 36  & 132  & 8.08:& \\ 
\qq& HS 2259+2357   & 23 01 38.0 & +24 14 05 &  7791 & $\pm$ 12 & 18.04 & 0.18 & --17.25 & 43  & 101  & 8.00:& \\
\qq& HS 2304+2255   & 23 07 18.9 & +23 11 54 &  6293 & $\pm$ 10 & 17.75 & 0.27 & --17.17 & 114 & 700  & 8.03:& \\
\qq& HS 2311+2631   & 23 13 48.0 & +26 48 02 &  6008 & $\pm$ 60 & 18.34 & 0.20 & --16.41 & 39  & 237  & 8.08 & \\
\qq& HS 2336+1800   & 23 38 46.8 & +18 17 20 &  4991 & $\pm$ 12 & 17.48 & 0.07 & --16.72 & 58  & 193  & 7.70 & \\
\qq& HS 2340+2034   & 23 42 38.2 & +20 51 31 &  2260 & $\pm$ 21 & 17.16 & 0.14 & --15.42 & 108 & 744  & 8.01 & \\
\qq& HS 2347+1618   & 23 50 06.0 & +16 35 06 & 14731 & $\pm$ 30 & 18.29 & 0.08 & --18.26 & 109 & 655  & 7.75 & \\
\qq& HS 2348+2920   & 23 50 50.4 & +29 37 03 &  4822 & $\pm$ 17 & 17.91 & 0.18 & --16.35 & 47  & 194  & 7.76 & \\
\qq& HS 2352+2733   & 23 54 56.7 & +27 49 59 &  8285 & $\pm$ 12 & 18.29 & 0.09 & --17.04 & 53  & 276  & 8.40:& \\
\qq& HS 2359+1659   & 00 02 09.9 & +17 15 59 &  6275 & $\pm$ 30 & 16.59 & 0.06 & --18.09 & 231 & 1537 & 8.09 & \\
\\[-0.25cm] \hline \\[-0.2cm]
\multicolumn{13}{l}{$^a$ Heliocentric velocities; $^b$ corrected APM magnitudes (Kniazev et al. 2002);
 $^c$ extinction estimates from NED following Schlegel et al.~(\cite{Schlegel98});} \\
\multicolumn{13}{l}{ $^d$ absolute magnitudes are corrected for galactic extinction;
 $^e$ less confident values are denoted by a colon } \\
\multicolumn{11}{l}{ References: {\bf 1} -- Popescu et al.~\cite{Popescu96}; {\bf 2} -- Lu et al.~\cite{Lu93}; {\bf 3} -- van Zee et al.~\cite{Zee96} } \\
\end{tabular}                                                                                                                                                                                                      
\end{center}
\end{table*}                                                                                                                                                                                                       


\end{landscape}
\clearpage

\clearpage

\begin{landscape}
\headsep 6cm
\renewcommand{\baselinestretch}{0.8}


\setcounter{qub}{0}

\begin{table*}[h]

\vspace{5.5cm}
\begin{center}

\caption{\label{Tab3} Parameters of Emission--Line Galaxies
			with EW$_{\lambda 5007}$ $<$ 100\AA.}

\vspace{-0.3cm}

\begin{tabular}{rlccrlcccccl} \hline \\[-0.2cm]
\multicolumn{1}{c}{\# }               &
\multicolumn{1}{c}{Name }             &
\multicolumn{1}{c}{$\alpha\,(2000)$ } &
\multicolumn{1}{c}{$\delta\,(2000)$ } &
\multicolumn{1}{c}{V$_\mathrm{hel}^{\;\;\;\;a}$ }  &
\multicolumn{1}{c}{$\pm\sigma_\mathrm{V}$ }  &
\multicolumn{1}{c}{m$_{B}^{\;\;b}$ }  &
\multicolumn{1}{c}{$A_{B}^{\;\;c}$ }  &
\multicolumn{1}{c}{M$_{B}^{\;\;d}$ }  &
\multicolumn{1}{c}{EW$_\mathrm{H\beta}$ }       &
\multicolumn{1}{c}{EW$_{\lambda 5007}$ } &
\multicolumn{1}{l}{Other names from NED } \\
&
\multicolumn{1}{c}{ }  &
\multicolumn{1}{c}{ }  &
\multicolumn{1}{c}{ }  &
\multicolumn{1}{c}{(km s$^{-1}$)}  &
\multicolumn{1}{c}{(km s$^{-1}$)}  &
\multicolumn{1}{c}{ }  &
\multicolumn{1}{r}{ }  &
\multicolumn{1}{r}{ }  &
\multicolumn{1}{c}{(\AA) }  &
\multicolumn{1}{c}{(\AA) }  &
\multicolumn{1}{l}{and number of reference }  \\
&
\multicolumn{1}{c}{ (1) }  &
\multicolumn{1}{c}{ (2) }  &
\multicolumn{1}{c}{ (3) }  &
\multicolumn{1}{c}{ (4) }  &
\multicolumn{1}{c}{ (5) }  &
\multicolumn{1}{c}{ (6) }  &
\multicolumn{1}{c}{ (7) }  &
\multicolumn{1}{c}{ (8) }  &
\multicolumn{1}{c}{ (9) }  &
\multicolumn{1}{c}{ (10)}  &
\multicolumn{1}{c}{ (11)}  \\
\\[-0.2cm] \hline \\[-0.2cm]
\qq& HS 0001+1703   & 00 04 11.7 & +17 19 49 &  5478 & $\pm$  99 & 17.79 & 0.07 & --16.61 & 5   & 31 & \\
\qq& HS 0053+2910   & 00 56 27.6 & +29 27 10 & 11041 & $\pm$ 100 & 17.00 & 0.20 & --19.05 & --~ & -- & NPM1G +29.0035, \\
   &                &            &           &       &           &       &      &         &     &    & 2MASXi J0056275+29271 \\
\qq& HS 2300+2612   & 23 03 20.2 & +26 28 52 & 23803 & $\pm$  87 & 17.20 & 0.20 & --20.52 & 17  & 25 & \\
\qq& HS 2319+2723   & 23 21 41.9 & +27 39 48 &  5936 & $\pm$  69 & 17.86 & 0.25 & --16.92 & --~ & 22 & ESDO F535-09, LEDA 141110, \\
   &                &            &           &       &           &       &      &         &     &    & [HR89] 231913.4+272316, {\bf 4} \\
\qq& HS 2336+2112   & 23 38 32.2 & +21 28 41 &  5619 & $\pm$ 111 & 17.05 & 0.09 & --17.43 & 16  & 82 & \\
\\[-0.25cm] \hline \\[-0.15cm]
\multicolumn{12}{l}{$^a$ heliocentric velocities; $^b$ corrected APM magnitudes (Kniazev et al.~\cite{HSS_4});
 $^c$ extinction estimates from NED following Schlegel et al.~(\cite{Schlegel98});} \\
\multicolumn{12}{l}{$^d$ absolute magnitudes are corrected for galactic extinction} \\
\multicolumn{11}{l}{References: {\bf 4} -- Eder et al.~\cite{Eder89}} \\
\end{tabular}
\end{center}
\end{table*}


\end{landscape}

\clearpage

\begin{figure*}[h]
\appendix
\section{Spectra of emission-line galaxies from the HSS--LM}
\vspace*{-0.9cm}
\psfig{figure=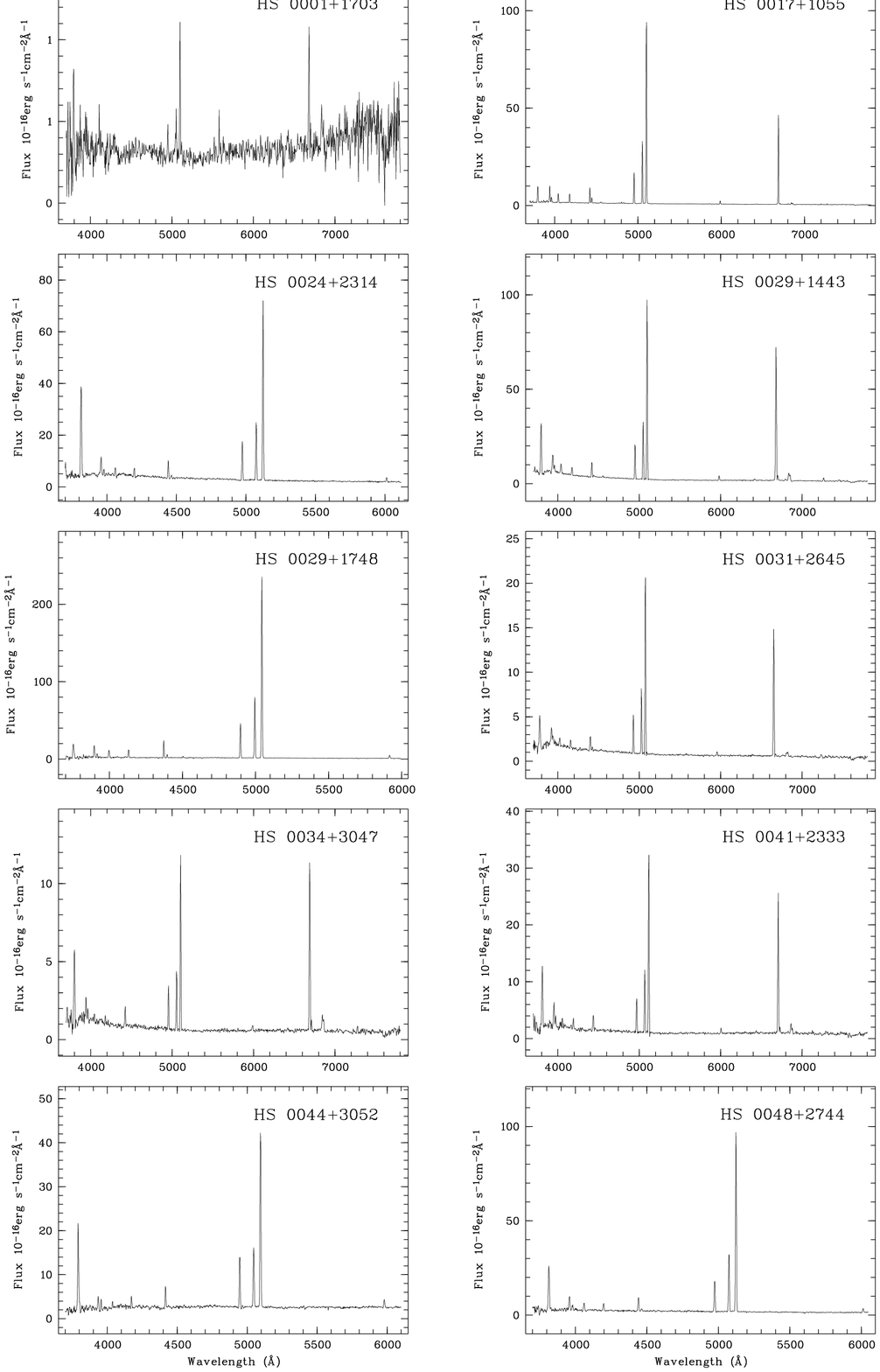,width=18.5cm,angle=0}
\vspace*{-1.0cm}
\centering
{\hspace*{0.5cm} \normalsize Fig. A1.}
\end{figure*}

\begin{figure*}[h]
\vspace*{-0.25cm}
\psfig{figure=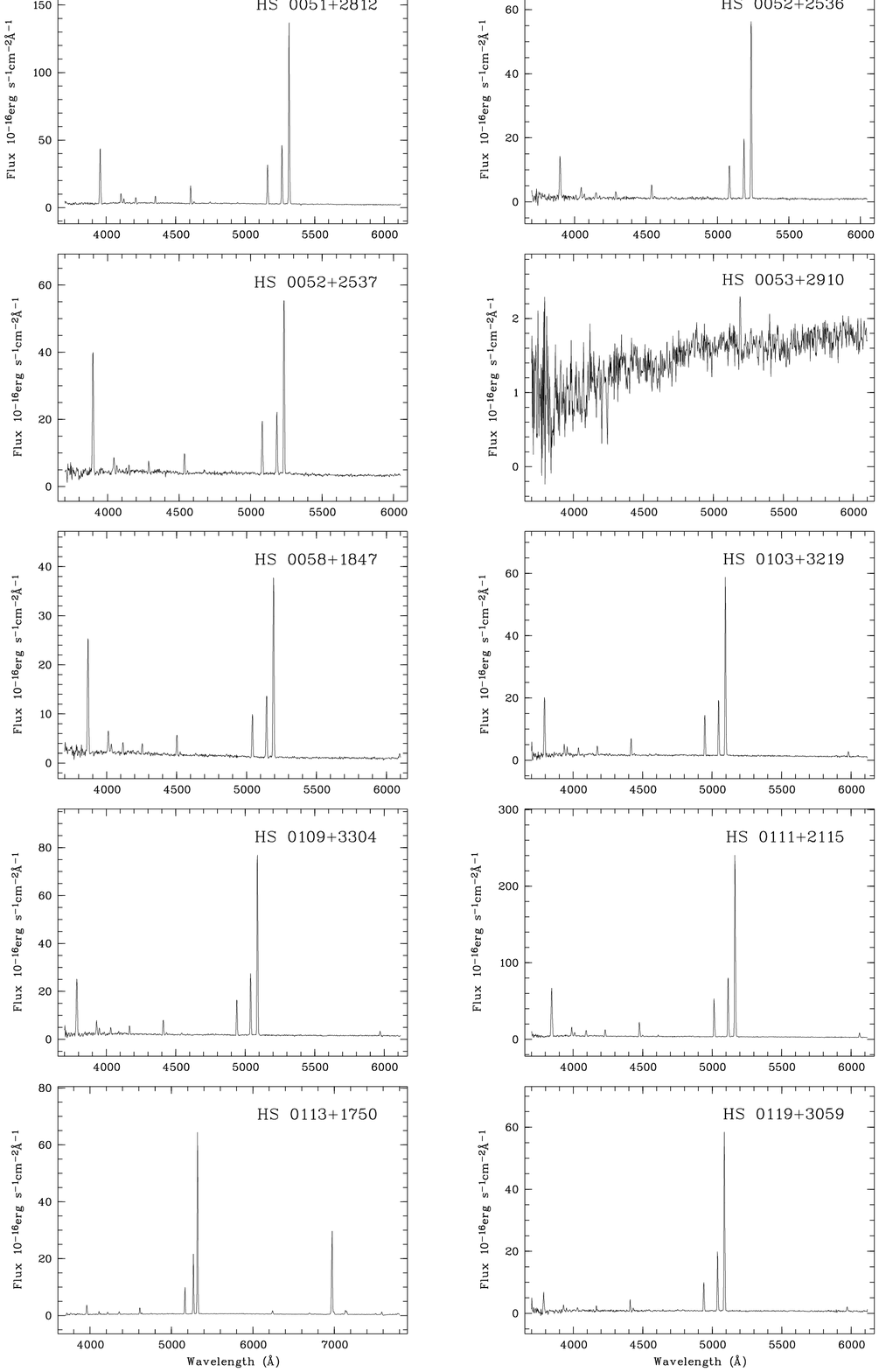,width=18.5cm,angle=0}
\vspace*{-1.0cm}
\centering
{\hspace*{0.5cm} \normalsize Fig. A2.}
\end{figure*}

\begin{figure*}[h]
\vspace*{-0.25cm}
\psfig{figure=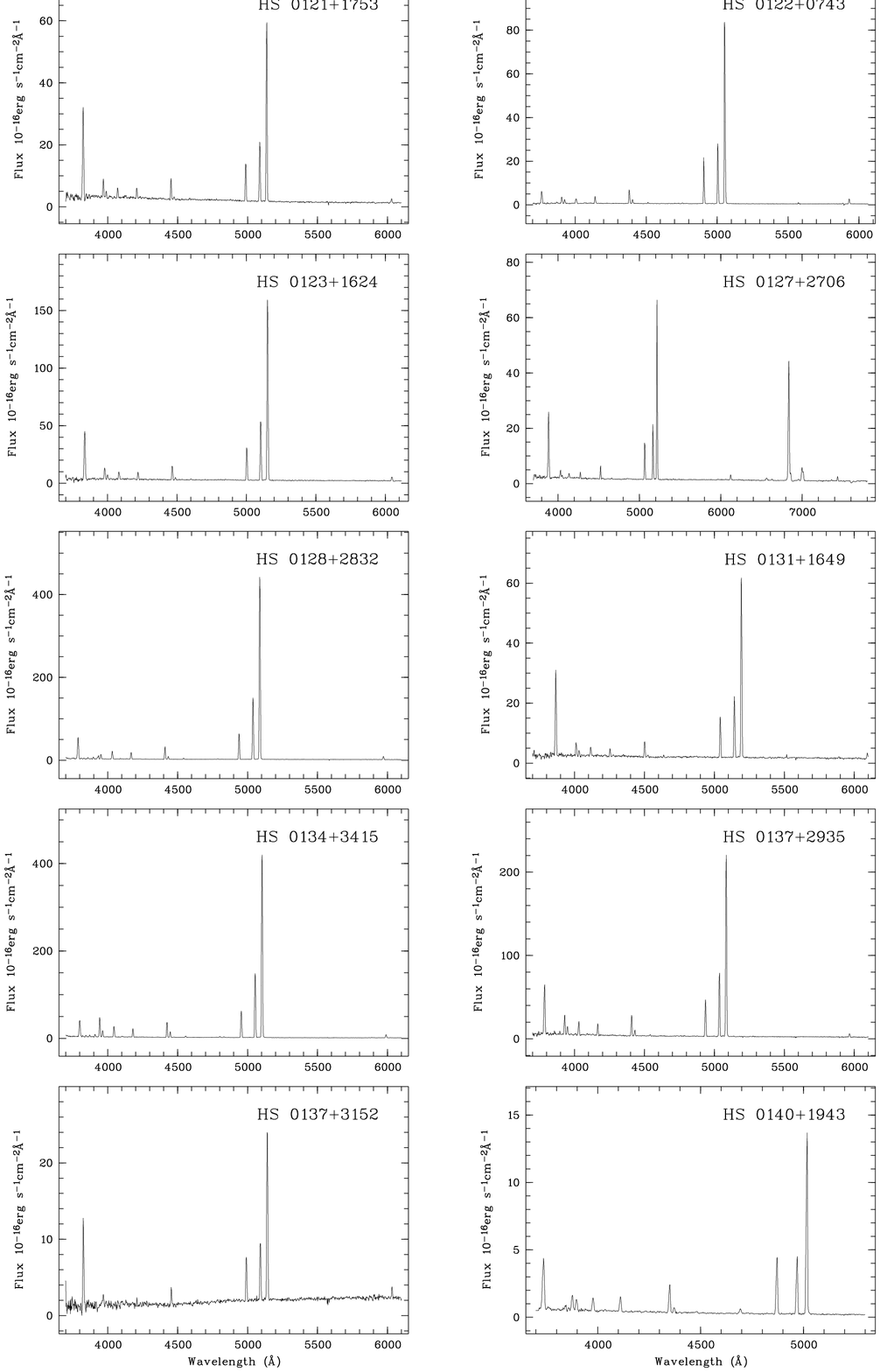,width=18.5cm,angle=0}
\vspace*{-1.0cm}
\centering
{\hspace*{0.5cm} \normalsize Fig. A3.}
\end{figure*}

\begin{figure*}[h]
\vspace*{-0.25cm}
\psfig{figure=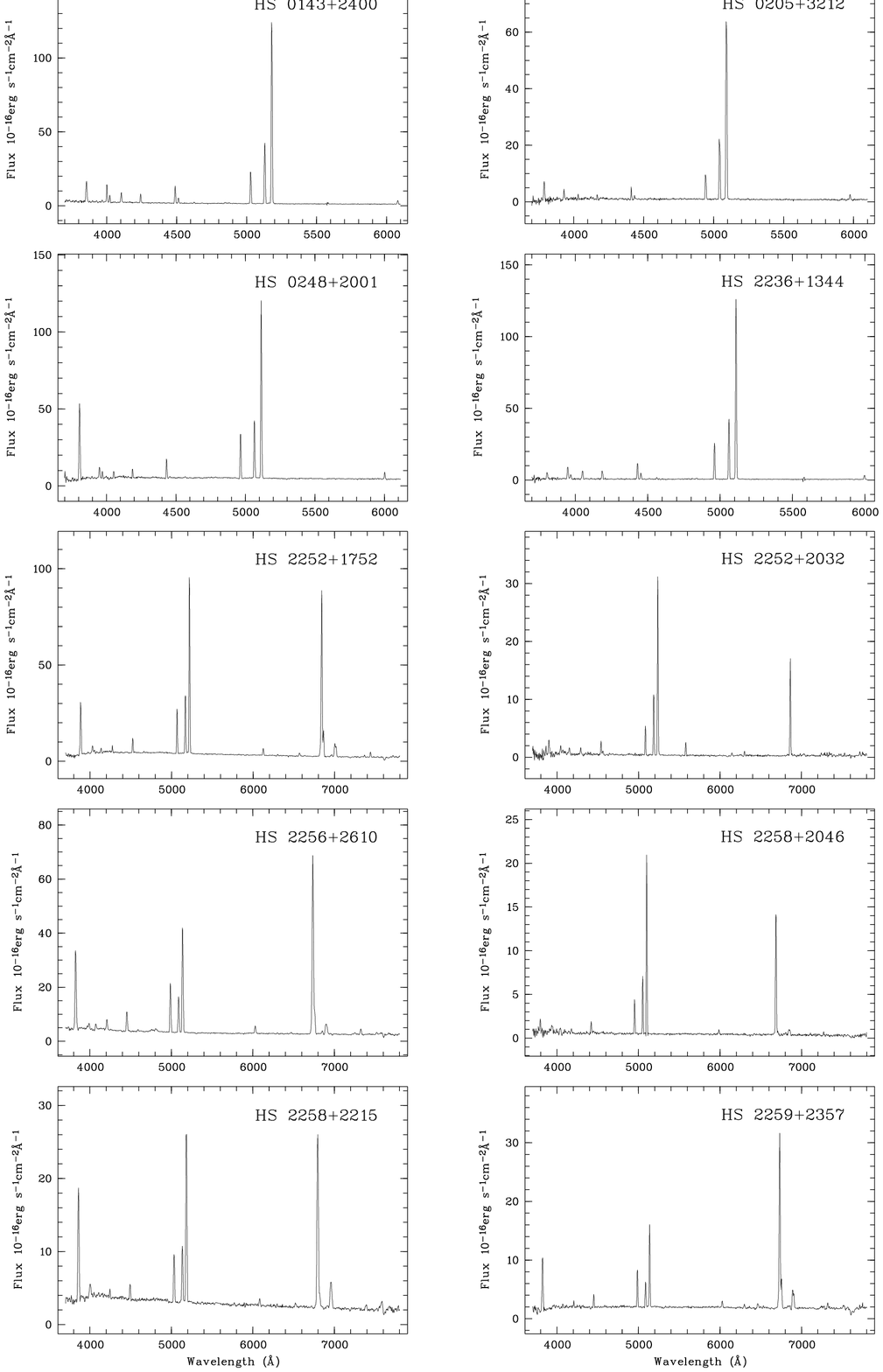,width=18.5cm,angle=0}
\vspace*{-1.0cm}
\centering
{\hspace*{0.5cm} \normalsize Fig. A4.}
\end{figure*}

\begin{figure*}[h]
\vspace*{-0.25cm}
\psfig{figure=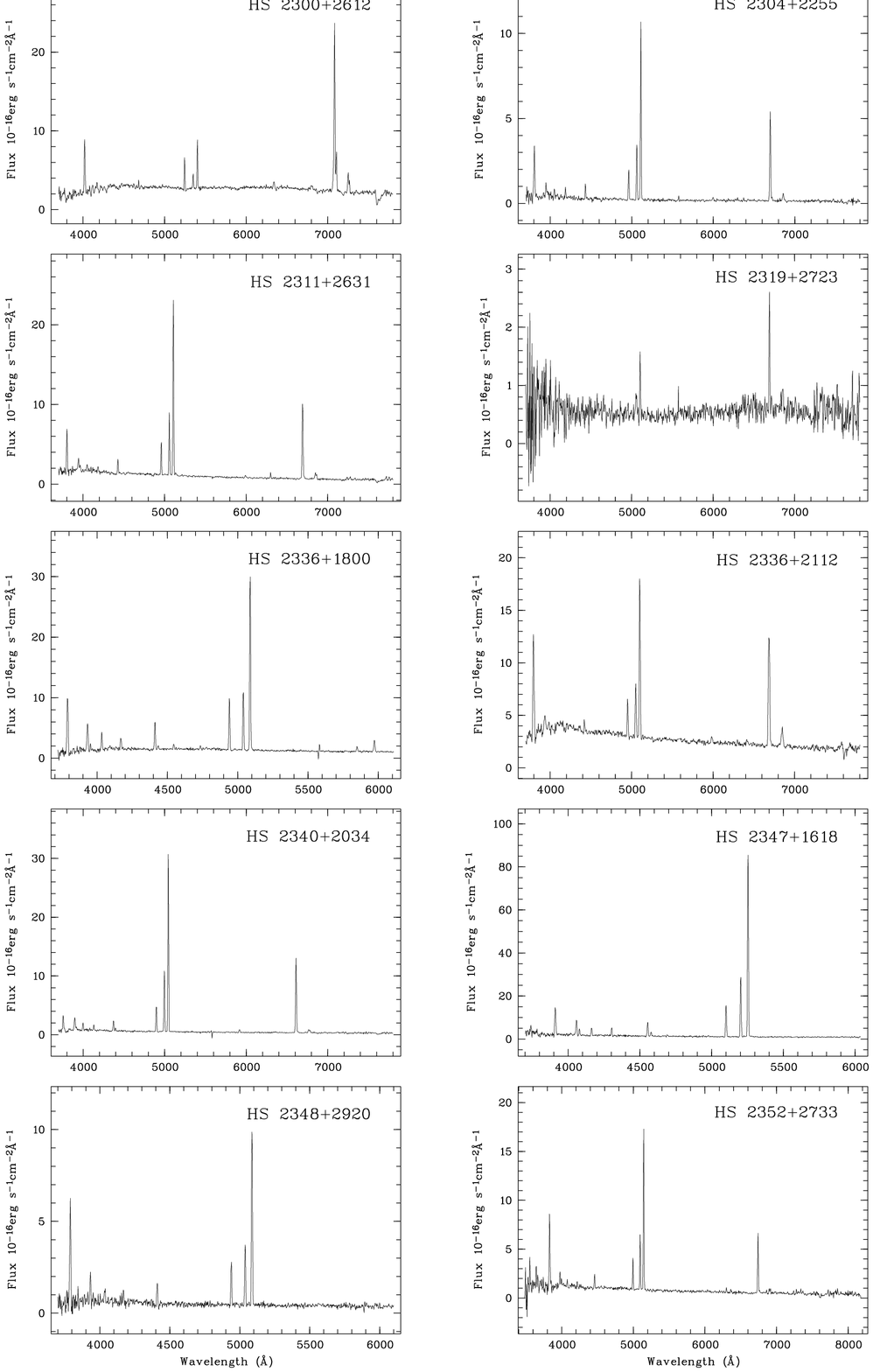,width=18.5cm,angle=0}
\vspace*{-1.0cm}
\centering
{\hspace*{0.5cm} \normalsize Fig. A5.}
\end{figure*}

\begin{figure*}[h]
\vspace*{-0.25cm}
\psfig{figure=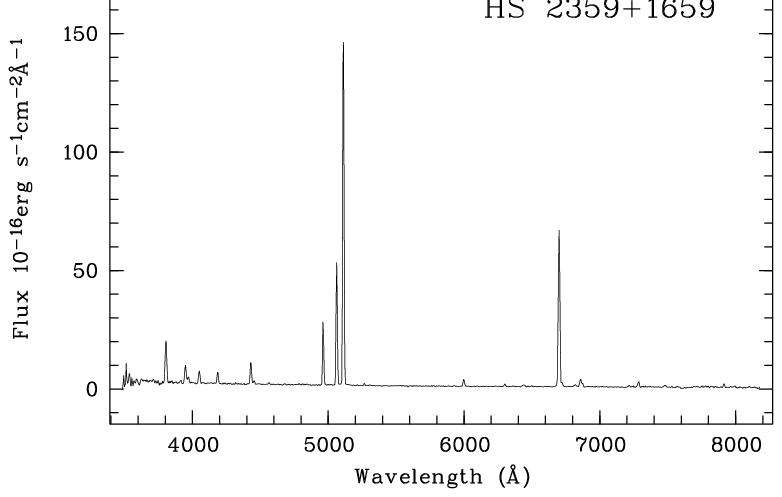,width=18.5cm,angle=0}
\vspace*{-1.0cm}
\centering
{\hspace*{0.5cm} \normalsize Fig. A6.}
\end{figure*}




\begin{thebibliography}{99}

\bibitem[1995]{Afanasiev95} Afanasiev, V. L., Burenkov, A. N., Vlasyuk, V.
  V., \& Drabek, S. V. 1995, SAO RAS internal report No.~234

\bibitem[1984]{Aller84} Aller, L. H. 1984, Physics of Thermal Gaseous Nebulae,
  Dordrecht: Reidel

\bibitem[1999]{Aloisi99} Aloisi, A., Tosi, M., \& Greggio, L. 1999, AJ, 118, 302

\bibitem[1999]{Bergvall99} Bergvall, N., R\"onnback, J., Masegosa, J.,
  \& \"Ostlin, G. 1999, A\&A, 341, 697

\bibitem[1996]{Bohlin96} Bohlin, R. C. 1996, AJ, 111, 1743

\bibitem[1993]{Campos93} Campos-Aguilar, A., Moles, M., \& Masegosa, J. 1993,
  AJ, 106, 1784

\bibitem[1997]{Doublier97} Doublier, V., Comte, G., Petrosian, A., et al.
  1997, A\&AS, 124, 405

\bibitem[1989]{Eder89} Eder, J. A., Oemler, A. J., Schombert, J. M., \&
 Dekel, A. 1989, ApJ, 340, 29


\bibitem[2001]{Fricke01} Fricke, K. J., Izotov, Y. I., Papaderos, P.,
  Guseva, N. G., \& Thuan, T. X. 2001, AJ, 121, 169

\bibitem[1995]{Hagen95} Hagen, H.-J., Groote, D., Engels, D., \& Reimers, D.
  1995, A\&AS, 111, 195

\bibitem[1956]{Haro56} Haro, G. 1956, Bol Obs. Tonantzintla y Tacubaya, 2, 8

\bibitem[1995]{Hopp95} Hopp, U., \& Kuhn, B. 1995, in Reviews in
  Modern Astronomy, ed. G. Klare, 8, 277

\bibitem[2000]{Hopp00} Hopp U., Engels D., Green, R., et al. 2000, A\&AS,
  142, 417  {\bf (HSS-III)}

\bibitem[1998]{Irwin98} Irwin, M. 1998,
  http://www.ast.cam.ac.uk/$^\sim$apmcat/

\bibitem[1999]{Izotov99} Izotov, Y. I., \& Thuan, T. X. 1999, ApJ, 511, 639

\bibitem[1993]{Izotov93} Izotov, Y. I., Lipovetsky, V. A., Guseva, N. G.,
  Kniazev A. Y., \& Stepanian, J. A. 1993, in DAEC Workshop, The Feedback of
  Chemical Evolution on the Stellar Content of Galaxies, eds. D. Alloin, \&
  G. Stasinska, Meudon Obs., 127

\bibitem[1994]{Izotov94} Izotov, Y. I., Thuan, T. X., \& Lipovetsky, V. A.
  1994, ApJ, 435, 647

\bibitem[1997]{Izotov97} Izotov, Y. I., Thuan, T. X., \& Lipovetsky, V. A.
  1997, ApJS, 108, 1

\bibitem[2002]{Izotov02} Izotov, Y. I., et al. 2002, in preparation

\bibitem[2001]{Izotov01} Izotov, Y. I., Chaffee, F. H., Foltz, C. B., et al.
  2001, ApJ, 560, 222

\bibitem[1995]{Kniazev95} Kniazev, A. Y., \& Shergin, V. S. 1995, SAO RAS
  internal report No.~249, 1

\bibitem[1998]{Kniazev98} Kniazev, A. Y., Pustilnik, S. A., \& Ugryumov, A. V.
   1998, Bulletin SAO, 46, 23

\bibitem[2000a]{Kniazev00a} Kniazev, A. Y., Pustilnik, S. A., Ugryumov, A. V.,
   \& Kniazeva, T. F., 2000a, Astronomy Lett., 26, 163

\bibitem[2000b]{Kniazev00b} Kniazev, A. Y., Pustilnik, S. A., Masegosa, J.,
  et al. 2000b, A\&A, 357, 101

\bibitem[2001]{HSS_4} Kniazev, A. Y., Engels D., Pustilnik, S. A., et al.
  2001, A\&A, 366, 771  {\bf (HSS-IV)}

\bibitem[2002]{Kniazev02} Kniazev, A. Y., Pustilnik, S. A., Ugryumov, A. V.,
  et al. 2002, in preparation

\bibitem[2000]{KO} Kunth, D., \& \"Ostlin, G. 2000, A\&AR, 10, 1

\bibitem[1988]{Kunth88} Kunth, D., Maurogordato, S., \& Vigroux, L. 1988,
  A\&A, 204, 10

\bibitem[2002]{Legrand02} Legrand, F., Tenorio-Tagle, G., Silich, S, Kunth,
  D., \& Servi\~no, M. 2002, AJ, in press = astro-ph/0106431

\bibitem[1999]{Starburst99} Leitherer, C., Schaerer, D., Goldader, J. D.,
  et al. 1999, ApJS, 123, 3

\bibitem[1999]{Lipovetsky99} Lipovetsky, V. A., Chaffee, F. H., Izotov, Y. I.,
  et al. 1999, ApJ, 519, 177

\bibitem[1993]{Lu93}  Lu, N. Y., Hoffman, G. L., Groff, T., Roos, T., \&
  Lamphier, C. 1993, ApJS, 88, 383

\bibitem[1977]{McAlpine77} MacAlpine, G. M., Smith, S. B., \& Lewis, D. W.
  1977, ApJS, 34, 95

\bibitem[1999]{MLF99} Mac Low, M.-M., \& Ferrara, A. 1999, ApJ, 513, 142

\bibitem[1999]{Mamon99} Mamon G. A. 1999, in ASP Conference Series
  Cosmic Flows: Towards an Understanding of the Large-Scale Structure of
  the Universe, eds. S. Courteau, M. Strauss, \& J. Willick, in press

\bibitem[1967]{Markarian67} Markarian, B. E. 1967, Afz, 3, 55

\bibitem[1983]{Markarian83} Markarian, B. E., Lipovetsky, V. A., \& Stepanian,
  J. A. 1983, Afz, 19, 29

\bibitem[1994]{Masegosa94} Masegosa, J., Moles, M., \& Campos-Aguilar, A.
  1994, ApJ, 420, 576

\bibitem[2000]{Mouri00} Mouri, H., \& Taniguchi, Y. 2000, ApJ, 545, L103

\bibitem[1990]{Oke90} Oke, J. B. 1990, AJ, 99, 1621

\bibitem[2000]{Ostlin00} \"Ostlin, G., 2000, ApJ, 535, L99

\bibitem[2001]{OK01} \"Ostlin, G., \& Kunth, D. 2001, A\&A, 371, 429


\bibitem[1992]{Pagel92} Pagel, B. E. J., Simonson, E. A., Terlevich, R. J.,
  \& Edmunds, M. G. 1992, MNRAS, 255, 325

\bibitem[1998]{Papaderos98} Papaderos, P., Izotov, Y. I., Fricke, K. J.,
  Thuan, T. X., \& Guseva, N. G. 1998, A\&A, 338, 43

\bibitem[2002]{Papaderos02} Papaderos, P., Izotov, Y.I.,  Thuan, T.X., et al.
   2002, A\&A, 393, 461

\bibitem[1995]{Pesch95} Pesch, P., Stephenson, C. B., \& MacConnell, D. J.
  1995, ApJS, 98, 41

\bibitem[2000]{Pilyugin00} Pilyugin, L. 2000, A\&A, 362, 325

\bibitem[2001]{Pilyugin01} Pilyugin, L. 2001, A\&A, 369, 594

\bibitem[1996]{Popescu96} Popescu, C. C., Hopp, U., Hagen, H.-J., \&
  Els\"{a}sser, H. 1996, A\&AS, 116, 43

\bibitem[1998]{Popescu98} Popescu, C. C., Hopp, U., Hagen, H.-J.,
  \& Els\"{a}sser, H. 1998, A\&AS, 133, 13

\bibitem[1995]{PULTG} Pustilnik, S. A., Ugryumov, A. V., Lipovetsky, V. A.,
  Thuan, T. X., \& Guseva, N. G. 1995, ApJ, 443, 499

\bibitem[1999]{Pustilnik99} Pustilnik, S. A., Engels, D., Ugryumov, A. V.,
  et al. 1999, A\&AS, 137, 299 {\bf(HSS-II)}

\bibitem[2001a]{PBTLI} Pustilnik, S. A., Brinks, E., Thuan, T. X.,
  Lipovetsky, V. A., \& Izotov, Y. I. 2001a, AJ, 121, 1413

\bibitem[2001b]{BCG_ENV} Pustilnik, S. A., Kniazev, A. Y., Lipovetsky, V. A.,
  \& Ugryumov, A. V. 2001b, A\&A, 373, 24

\bibitem[2002a]{PKM02} Pustilnik, S. A., Kniazev, A. Y., Masegosa, J., et al.
   2002a, A\&A, 389, 779

\bibitem[2002b]{LSBD} Pustilnik, S. A., Kniazev, A. Y., Pramsky, A. G.,
  Ugryumov, A. V., \& Masegosa, J. 2002b, A\&A, submitted

\bibitem[2002c]{PEM02} Pustilnik, S. A., Engels, D., Masegosa, J., et al.
  2002c, A\&A, in preparation {\bf(HSS-VI)}

\bibitem[1995]{RM95} Richer, M. G., \& McCall, M. L. 1995, ApJ, 445, 642

\bibitem[1989a]{Salzer89a} Salzer, J. J. 1989a, ApJ, 347, 152

\bibitem[1999]{SN99} Salzer, J. J., \& Norton, S. A. 1999, in ASP Conference
  Series 170, Proc. of IAU colloq., Low Surface Brightness Universe,
  eds. J. I. Davies, C. Impey, \& S. Phillipps, 253

\bibitem[1989b]{Salzer89b} Salzer, J. J., McAlpine, G. M., \& Boroson T. A.
  1989b, ApJS, 70, 447

\bibitem[1991]{Salzer91} Salzer, J. J.,  Di Serego Alighieri, S., Matteucci,
   F., Giovanelli, R., \& Haynes, M. P. 1991, AJ, 101, 1258

\bibitem[1995]{Salzer95} Salzer, J. J., Moody, J. W., Rosenberg, J. L.,
  Gregory, S. A., \& Newberry, M. V. 1995, AJ, 109, 2376

\bibitem[1970]{Sargent70} Sargent, W. L. W., \& Searle, L. 1970, ApJ,
  162, L155

\bibitem[1998]{SV98} Schaerer, D., \& Vacca, W. D. 1998, ApJ, 497, 618

\bibitem[1997]{Schaerer97} Schaerer, D., Contini, T., Kunth D., \& Meynet G.
  1997, ApJ, 481, L75

\bibitem[1999]{Schaerer99} Schaerer, D., Contini, T., \& Kunth D.
  1999, A\&A, 341, 399

\bibitem[1972]{Searle72} Searle, L., \& Sargent, W. L. W. 1972, ApJ, 173, 25

\bibitem[1998]{Schlegel98} Schlegel, D., Finkbeiner, D. P., \& Douglas, M.
  1998, ApJ, 500, 525


\bibitem[2001]{STT01} Silich, S., \& Tenorio-Tagle, G. 2001, ApJ, 552, 91

\bibitem[1988]{Skillman88} Skillman, E., Melnick, J., Terlevich, R., \&
  Moles, M. 1988, A\&A, 196, 31

\bibitem[1989]{Skillman89} Skillman, E., Kennicutt, R., \& Hodge, P. 1989,
  ApJ, 347, 875

\bibitem[1996]{SL96}Stasinska, G., \& Leitherer, C. 1996, ApJS, 107, 661


\bibitem[1994]{Stepanian94} Stepanian, J. A. 1994, Proc. IAU Symp. 161,
  Kluwer, Dordrecht, eds. H. T. MacGillivray, et al., 731

\bibitem[1995]{Taylor95} Taylor, C. L., Brinks, E., Grashuis, R. M., \&
  Skillman, E. D. 1995, ApJS, 99, 427

\bibitem[1997]{Telles97} Telles, E., \& Terlevich, R. 1997, MNRAS, 286, 183

\bibitem[1991]{Terlevich91} Terlevich, R., Melnick, J., Masegosa, J., Moles,
  M., \& Copetti, M. V. F. 1991, A\&AS, 91, 285

\bibitem[1995]{Thuan95} Thuan, T. X., Izotov, Y. I., \& Lipovetsky, V. A.
  1995, ApJ, 445, 108

\bibitem[1999a]{TLMP99} Thuan, T. X., Lipovetsky, V. A., Martin, J.-M.,
  \& Pustilnik, S. A. 1999a, A\&AS, 139, 1

\bibitem[1999b]{Thuan99} Thuan, T. X., Izotov, Y. I., \& Foltz, C. B. 1999b,
  ApJ, 525, 105

\bibitem[1998]{Ugryumov98} Ugryumov, A. V., Pustilnik, S. A., Lipovetsky, V. A.,
  Richter, G. M., \& Izotov, Y. I. 1998, A\&AS, 131, 285

\bibitem[1999]{Ugryumov99} Ugryumov, A. V., Engels, D., Lipovetsky, V. A.,
  et al. 1999, A\&AS, 135, 511 {\bf(HSS-I)}

\bibitem[2001]{Ugryumov01} Ugryumov, A. V., Engels, D., Kniazev, A. Y.,
  et al. 2001, A\&A, 374, 907 {\bf(HSS-V)}

\bibitem[1995]{Vilchez95} Vilchez, J. M. 1995, AJ, 110, 1090

\bibitem[2000]{York2000} York, D. G., Adelman, J., Anderson, J. E., et al. 2000,
  AJ, 120, 1579

\bibitem[2000]{Zee00} van Zee, L., 2000, ApJ, 543, L31

\bibitem[1996]{Zee96} van Zee, L., Haynes, M. P., Salzer, J., \& Broelis, A.
  1996, AJ, 112, 129

\bibitem[1998a]{Zee98a} van Zee, L., Skillman, E. D., \& Salzer, J. J.
  1998a, AJ, 116, 1186

\bibitem[1998b]{Zee98b} van Zee, L., Westpfahl, D., Haynes, M., \& Salzer, J.
  1998b, AJ, 115, 1000

\bibitem[1998c]{Zee98c} van Zee, L., Salzer, J. J., Haynes, M. P., O'Donoghue,
  A. A., Balonek, T. J. 1998c, AJ, 116, 2805

\bibitem[1997]{Zenina97} Zenina, O. A., Balinskaya, I. S., Kniazev, A. Y., \&
  Lipovetsky, V. A. 1997, Astronomy Reports, 41, 472

\bibitem[1966]{Zwicky} Zwicky, F. 1966, ApJ, 143, 192
\end{thebibliography}
\end{document}